\documentclass[12pt,aps,prd,nofootinbib,floatfix,superscriptaddress,a4paper]{revtex4}

\usepackage{times,axodraw}
\usepackage[]{graphicx}
\usepackage{subfigure}
\usepackage{amssymb}
\usepackage{mathtools}

\def\lsim{\raise0.3ex\hbox{$\;<$\kern-0.75em\raise-1.1ex\hbox{$\sim\;$}}}
\def\gsim{\raise0.3ex\hbox{$\;>$\kern-0.75em\raise-1.1ex\hbox{$\sim\;$}}}
\def\321{SU(3) $\otimes$ SU(2) $\otimes$ U(1)}
\def\lr{ SU(2)$_L~\otimes$ SU(2)$_R~\otimes$ U(1)$_{B-L}$}

\begin{document}

\title{Lepton Flavour Violation and Flavour Symmetries}                       

\author{Frank~F.~Deppisch} \email{f.deppisch@ucl.ac.uk}         \affiliation{Department of Physics and Astronomy, University College London, United Kingdom}

\begin{abstract}\noindent
In this review article, we highlight the impact of models incorporating flavour symmetries on charged lepton flavour violating (LFV) processes. Flavour symmetries provide a natural approach to explain the peculiar mass hierarchies and mixing patterns of the Standard Model fermions. New sources of LFV are generally present in new physics beyond the Standard Model, and flavour symmetries can make distinctive predictions for LFV observables. Their discovery can provide crucial information to distinguish flavour symmetries and new physics scenarios in general. Because of their high sensitivity, we will focus on searches for low-energy LFV processes such as $\mu\to e\gamma$ or $\mu-e$ conversion in nuclei but we will also highlight the potential impact of LFV processes at the LHC. If new physics occurs at a scale accessible by the LHC, the flavour operators they induce could potentially be probed in high detail. 
\end{abstract}

\maketitle

\section{Introduction}
\label{sec:introduction}

The flavour puzzle is one of the main unresolved problems of particle physics: So far we do not understand why there are three generations of fermions and we have no firm explanation for their peculiar structure of mass hierarchies and mixing properties. Experimentally, many of the properties are already well known, such as the strong hierarchy of quark masses,
\begin{equation}
	m_u : m_c : m_t \sim \lambda^8 : \lambda^4 : 1, \quad
	m_d : m_s : m_b \sim \lambda^4 : \lambda^2 : 1, \quad 
	\lambda \sim 0.2,
\end{equation}
whereas the mixing between quarks is found to be weak, with the small Kobayashi-Maskawa mixing angles
\begin{equation}
	\theta^Q_{12} \sim 13^\circ, \quad
	\theta^Q_{23} \sim 2.4^\circ, \quad
	\theta^Q_{13} \sim 0.2^\circ.
\end{equation}
The hierarchy of the charged lepton masses is similar to the hierarchy of the down-type quarks, and the neutrino mass differences derived in oscillation experiments~\cite{fukuda:1998mi, ahmad:2002jz, eguchi:2002dm} as well as limits on the absolute neutrino mass scale from neutrinoless double beta decay, Tritium decay and cosmological observations suggest a much more shallow hierarchy for the neutrinos,
\begin{align}
	m_e : m_\mu : m_\tau        &\sim     \lambda^{4.5} : \lambda^2 : 1,  \nonumber\\
	m^2_{\nu_2} - m^2_{\nu_1}   &\sim     7.6 \times 10^{-5}\text{ eV}^2, \nonumber\\
	|m^2_{\nu_3} - m^2_{\nu_1}| &\sim     2.4 \times 10^{-3}\text{ eV}^2, \nonumber\\
	\min(m_{\nu_i})             &\lesssim 0.3\text{eV},               
\end{align}
i.e. even in the extreme case of a massless lightest neutrino, the hierarchy of the other two neutrinos would be only $m_{\nu_3} : m_{\nu_2} \sim \lambda$. The leptonic mixing, also derived from neutrino oscillations, exhibits a pattern that is substantially different from that which characterises quarks~\cite{nunokawa:2007qh},
\begin{equation}
	\theta^L_{12} \sim 30^\circ, \quad
	\theta^L_{23} \sim 45^\circ, \quad
	\theta^L_{13} \sim 8^\circ.
\end{equation}

One of the main goals of physics beyond the Standard Model (BSM) is to solve this flavour puzzle. Given the success of gauge symmetries to describe the structure of particle interactions, it is natural to try to explain the fermion mass and mixing structure in terms of a symmetry in the space of the fermion generations, i.e. a flavour symmetry. The appearance of three fermion generations can be understood as a U(3)$^5$ symmetry in the flavour space of the 5 fermion species $Q_i, u_i^c, d_i^c, L_i, e_i^c$, $i=1,2,3$. In the absence of Yukawa interactions, the SM Lagrangian is invariant under such a U(3)$^5$ transformation. This invariance is broken by the Yukawa interactions and the question is in what form this breaking occurs.

An explanation of the observed patterns can made in different directions. In Grand Unified Theories (GUTs), the different SM fermion families are unified, and their Yukawa interactions are no longer independent from each other. For example in SO(10), one generation of fermions, including the right-handed neutrino, unify in one representation. This reduces the maximal flavour symmetry to U(3). On the other hand, flavour symmetries try to explain the pattern of Yukawa interactions in terms of a spontaneous or explicit breaking of $U(3)^5$. There are many options to implement such a breaking, for example depending on whether the symmetry under consideration is continuous or discrete, Abelian or non-Abelian, and whether it is broken at low energy scale accessible in experiments or at a high energy scale. The latter is usually preferred to avoid unwanted phenomenological implications of new particles and operators, such as additional contributions to flavour changing neutral currents (FCNCs).

Despite the wealth of experimental data on the structure and couplings of the fermion sector, more information is required to unravel the degeneracies in the predictions of the various models. A highly important field of phenomenology in this regard are charged lepton flavour violating (LFV) processes, as they provide crucial information on the flavour structure of the leptonic sector in many theories of BSM physics.
\begin{figure}[t]
\centering
\includegraphics[clip,width=0.40\textwidth]{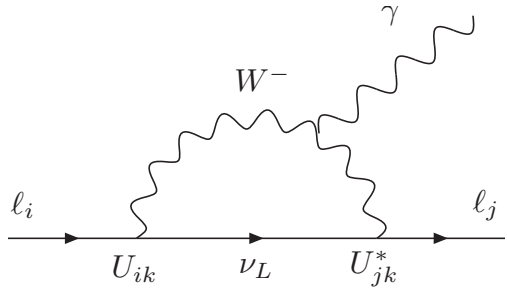}
\caption{Radiative decays $\ell_i \to \ell_j \gamma$ in the SM with massive light neutrinos (from \cite{Abada:2011rg}).}
\label{fig:LFV}
\end{figure}
The discovery of neutrino oscillations~\cite{fukuda:1998mi, ahmad:2002jz, eguchi:2002dm} shows that neutrinos are massive~\cite{Maltoni:2004ei} and that lepton flavour is violated in neutrino propagation. It is natural to expect that the violation of this conservation law should show up in other contexts, such as rare lepton flavour violating (LFV) decays of muons and taus, e.g. $\mu^-\to e^-\gamma$ and possibly also at the high energies accessible at the LHC. Unfortunately, in the SM with light left-handed neutrinos, LFV is naturally suppressed via the GIM mechanism due to the small neutrino masses. For example, the LFV process $\mu\to e\gamma$ proceeds via the diagram in Fig.~\ref{fig:LFV}, and its branching ratio is given by ($U$ is leptonic PMNS mixing matrix)
\begin{equation}
\label{eq:BrMuEGamma_SM}
	Br(\mu\to e\gamma) = \frac{3 \alpha}{32 \pi} \left|   
	\sum_i U^*_{\mu i} U_{e i}\frac{m^2_{\nu_i}}{m_W^2}
	\right|^2 \lesssim 10^{-54},
\end{equation}
far below the experimental sensitivity in the foreseeable future. Analogous results hold for all other LFV processes. On the other hand, this means that the observation of charged lepton flavour violation would be a clear signal of new physics. In general, physics beyond the Standard Model contains new sources of flavour violation, typically induced by breaking lepton flavour at a high scale. For example, the massive neutrinos in the supersymmetric Seesaw mechanism affect the renormalization group equations of the slepton masses and the trilinear couplings, and give rise to non-diagonal matrix elements inducing LFV. The non-observation of associated LFV processes already provides strong bounds on BSM physics.
In the context of flavour symmetries, different symmetries and breaking patterns can be distinguished by the relative strength of LFV processes they predict, such as $Br(\mu\to e\gamma) : Br(\tau\to e\gamma): Br(\tau\to\mu\gamma)$. In addition, there is also an interplay with the general BSM framework into which a flavour symmetry is embedded, such as supersymmetry, which determines the mass scale that induces new physics effects. 

In this review article, we will highlight the impact of lepton flavour violating processes on models incorporating flavour symmetries. Because of their high sensitivity, searches for low-energy LFV processes such as $\mu\to e\gamma$ or $\mu-e$ conversion in nuclei will be a strong focus here, but we will also highlight the potential impact of LFV processes at the LHC. If new physics occurs at a scale accessible at the LHC, as is generally expected in order to understand electroweak symmetry breaking, the flavour operators they induce could potentially be probed in high detail. 

The article is organised as follows. In Section~\ref{sec:symmetries}, we provide a basic introduction to flavour symmetries as far as required for the article. A more thorough discussion of the theoretical foundations of flavour symmetries can be found elsewhere in this special issue. Section~\ref{sec:lfv} gives an overview of the experimental status of LFV searches as well as their future prospects, and in Section~\ref{sec:models} we discuss i) a model-independent approach to analyse the impact of flavour symmetries on the strength of LFV operators, as well as ii) several specific new physics models. The list of models and symmetries discussed here is by far not exhaustive, but is intended to highlight the interplay between the choice of flavour symmetry, the embedding BSM framework and the experimental LFV observables. We conclude this article with a summary and an outlook in Section~\ref{sec:summary}.

\section{Flavour Symmetries}
\label{sec:symmetries}

As discussed above, the Standard Model without Yukawa couplings is invariant under U(3)$^5$, with each term U(3) operating on the 3-dimensional generational space of one of the 5 fermion families. Models with flavour symmetries describe the generation of an appropriate texture of Yukawa interactions arising from the spontaneous or explicit breaking of a subgroup of U(3)$^5$. In addition, the broken symmetry can be either local or global, and may commute with the gauge symmetry or not. In any case one has to be careful to avoid unwanted effects from new particles or couplings associated with the breaking of the flavour symmetry, such as additional gauge interactions or pseudo-Goldstone bosons. In discrete symmetries this can be avoided, but the breaking of such symmetries generally leads to the creation of domain walls. To avoid such dangerous effects it is assumed in most flavour symmetry models that the breaking occurs via a vacuum expectation value (VEV) of one or more SM gauge singlets (flavons) at a high energy scale such as the GUT scale. The breaking may also occur at a scale close to the electroweak scale, with additional SU(2) Higgs doublets present (see e.g. \cite{Bhattacharyya:2010hp}). Such models can generate flavour changing neutral currents at the tree level, and they have to be carefully designed to prevent large LFV rates by flavour alignment or an appropriate choice of the flavour symmetry \cite{Joshipura:1990pi, Branco:1996bq, Kubo:2003iw, Botella:2009pq}.

In the generally assumed picture of a flavour symmetry breaking at a high scale, one or more flavons charged under this symmetry acquire a VEV. This is mediated to the SM fields via a messenger sector at a scale $M_F$ somewhat higher than the flavour symmetry breaking scale. The details of the breaking of the flavour symmetry and messenger sector are often not specified, but taken into account using an effective operator approach by integrating out the messenger fields below $M_F$. The generated Yukawa interactions are then described as expansions of non-renormalizable operators between the flavons and the fields participating in Yukawa interactions.

There is a huge variety of possible groups and breaking patterns that can be used to describe a flavour symmetry. In the following we will briefly highlight two examples of continuous and discrete groups to establish the basic idea of flavour symmetries. A more in-depth introduction to the theoretical framework can be found elsewhere in this special issue.

\subsection{Froggatt-Nielsen Mechanism}
\label{sec:FroggatNielsen}

The most famous example for a continuous flavour symmetry is the model of Froggatt-Nielsen~\cite{Froggatt:1978nt}, where the SM fermion generations are assigned different charges under a flavour group $U(1)_F$. The SM fields couple via a heavy mediator to a flavon field $\theta$ with flavour charge $-1$. When the flavon field acquires a vacuum expectation value (VEV) $v_F$, which is assumed to be smaller than the mass of mediator $M_F$, Yukawa interactions such as $Y^e_{ij} L_i e^c_j H_d$ ($i,j=1,2,3$) are generated\footnote{We implicitly work in a SUSY environment with two Higgs doublets $H_u$ and $H_d$ generating the fermion masses.}, which are suppressed by $\epsilon \equiv v_F/M_F \ll 1$ to the power of the sum of the $U(1)_F$ charges of the SM fields, $Y^e_{ij} = c^e_{ij} \epsilon^{Q^L_i + Q^{e^c}_j + Q^{H_d}}$. Here, $c^e_{ij}$ are the coupling coefficients of the underlying flavour interaction which can not determined within the framework but are assumed to be of order one. If for example one chooses the charges $Q^L_{e,\mu\tau} = (0,2,3)$, $Q^{e^c}_{e,\mu\tau} = (0,0,1)$, $Q^{H_d} = 0$, the charged lepton Yukawa coupling matrix is given by
\begin{equation}
	Y^e \sim
	\begin{pmatrix} 
		\epsilon^4 & \epsilon^3 & \epsilon^3 \\
		\epsilon^3 & \epsilon^2 & \epsilon^2 \\ 
		\epsilon   &          1 &          1 
	\end{pmatrix},
\end{equation}
where the entries are implicitly assumed to contain coefficients $c^e_{ij} = \mathcal{O}(1)$. Analogously, the structure of the quark Yukawa couplings and the neutrino mass matrix (via the effective operator $(\bar L_i\cdot H_u)(H_u\cdot L_j)^c/M_{B-L}$) can be modeled, through an appropriate choice of flavour charges. Fitting the observed fermion mass hierarchy suggests the order of the Cabibbo angle for $\epsilon \approx \lambda \approx 0.2$.

A large number of modifications and adaptations of this idea are possible, by using non-Abelian continuous symmetries such as U(2) \cite{Barbieri:1996ae, Barbieri:1996ww, Barbieri:1997tu}, SO(3) \cite{King:2005bj} or SU(3) \cite{King:2001uz, King:2003rf}, with sophisticated breaking patterns and the inclusions of additional symmetries like $U(1)$ or $Z_n$ to forbid certain couplings in order to explain or predict (in the case of neutrinos) the fermion mass hierarchies and mixing properties.

\subsection{The group $A_4$}
\label{sec:A4}

One of the most popular discrete groups used to construct a flavour symmetry is $A_4$. The importance of this group for flavour model building stems from the fact that it predicts the quark mass matrices to transform identically, but at the same time containing a large mis-alignment between the charged lepton and neutrino sectors. To lowest order, this would explain the lack of mixing in the quark sector and the large mixing of tri-bimaximal form \cite{Harrison:2002er,He:2003rm} in the lepton sector. $A_4$ is the symmetry group of a regular tetrahedron, or equivalently, the group of even permutations of 4 objects. The group therefore has 4!/2 = 12 elements and contains three one-dimensional ($1$, $1'$, $1''$) and one three-dimensional ($3$) irreducible unitary representations. The group is generated by two operations $S$ and $T$, which can be represented by the permutations $S: (1234) \to (4321)$ and $T: (1234) \to (2314)$. These generators therefore satisfy
\begin{equation}
	S^2 = T^3 = (ST)^3 = 1.
\end{equation} 
The three singlets can then be generated by assigning  
\begin{align}
	  1:      S &= 1, T = 1,                        \\
	  1':     S &= 1, T = e^{2\pi i/3} = \omega,    \\
	  1'':    S &= 1, T = e^{4\pi i/3} = \omega^2,
\end{align}
whereas a triplet can be generated from
\begin{equation}
	S = \frac{1}{3}
	\begin{pmatrix}
		-1 &  2 &  2 \\
		 2 & -1 &  2 \\
		 2 &  2 & -1
	\end{pmatrix}, \quad
	T =
	\begin{pmatrix}
		 1 & 0        &  0     \\
		 0 & \omega^2 &  0     \\
		 0 & 0        & \omega
	\end{pmatrix}.
\end{equation}
In the $A_4$ flavour model introduced in \cite{Altarelli:2005yx}, the left-handed lepton doublets $L_i$ collectively transform as an $A_4$ triplet whereas the three right-handed gauge singlet leptons $e^c$, $\mu^c$ and $\tau^c$ are assigned to the three inequivalent $A_4$ singlets $1$, $1''$ and $1'$, respectively. The Higgs gauge doublets $H_u$ and $H_d$ (the model is formulated in a supersymmetric framework) do not transform under $A_4$. The flavon sector of the model contains two triplets $\varphi_S$, $\varphi_T$ and two singlets $\xi$, $\tilde\xi$. In order to explain the hierarchy of charged lepton masses, an additional Froggatt-Nielsen U(1)$_\text{FN}$ with associated flavon $\theta_\text{FN}$ is introduced, as outlined in Section~\ref{sec:FroggatNielsen}, under which the lepton singlets are charged as $Q^{e^c}_{e,\mu,\tau} = (0,2,4)$\footnote{The model presented in \cite{Altarelli:2005yx} also contains an additional $Z_3$ symmetry to remove unwanted terms which are allowed in $A_4$ by interchanging $\varphi_T \leftrightarrow (\varphi_S, \xi, \tilde\xi)$.}. The flavour symmetry breaking occurs close to a cut-off scale $M_F$ with the vacuum alignment 
\begin{equation}
	\langle\varphi_S\rangle        =    v_S (1,1,1), \quad
	\langle\varphi_T\rangle        =    v_T (1,0,0), \quad
	\langle\xi      \rangle        \neq 0, \quad
	\langle\tilde\xi\rangle        =    0, \quad
	\langle\theta_\text{FN}\rangle \neq 0.
\end{equation}
As a result, the mass matrix $m^l$ of charged leptons is diagonal,
\begin{equation}
	m^l = \langle H_d \rangle \frac{v_T}{M_F} 
	\text{diag}
	\left(
		c_e  \lambda^2,
		c_\mu\lambda,
		c_\tau
	\right),
\end{equation}
with the suppression $\lambda = \langle \theta_\text{FN} \rangle / M_F \approx 0.2$ and Froggatt-Nielsen coupling constants $c_{e,\mu,\tau} = \mathcal{O}(1)$ as before. Furthermore, the neutrino mass matrix $m^\nu$ generated by the Weinberg operator $(\bar L_i\cdot H_u)(H_u\cdot L_j)^c/M_{B-L}$ is given by (the product of two $A_4$ triplets transforms as $3 \times 3 = 1 + 1' + 1'' + 3 +3 $)
\begin{equation}
	\label{eq:MnuA4}
	m^\nu = \frac{\langle H_u \rangle^2}{M_{B-L}} 
	\begin{pmatrix}
		a + 2b/3 &  -b/3 &  -b/3 \\
		    -b/3 &  2b/3 & a-b/3 \\
		    -b/3 & a-b/3 &  2b/3
 	\end{pmatrix},
\end{equation}
with $a = c_a \langle \xi \rangle / M_F$, $b = c_b v_S / M_F$ ($c_{a,b} = \mathcal{O}(1)$). The matrix (\ref{eq:MnuA4}) leads to tri-bimaximal mixing, which is in stark contrast to the mixing of quarks generated in this model. This is because $v_S$ and $v_T$ break $A_4$ into two different subgroups each preserved in either the charged lepton and neutrino sector, whereas the breaking can be constructed such that the up and down quark sectors preserve the same subgroup, thus aligning the mass matrices. 

Of course, strict tri-bimaximal lepton mixing is excluded as it predicts a zero value for the oscillation angle $\theta_{13}$, whereas recent experimental results indicate $\sin^2 2\theta_{13} \approx 0.1$, significantly different from zero~\cite{An:2012eh, Ahn:2012nd}. In general, a deviation from tri-bimaximal form is expected when taking into account the higher-dimensional operators, although one has to be careful not to disturb the predictions for solar and atmospheric angles too much. Examples of $A_4$ models with large $\theta_{13}$ can be found in \cite{Adhikary:2008au, Ishimori:2012fg}.

The above procedure can be applied to a large number of different flavour symmetries, such as the permutation group $S_3$, and their breaking patterns. Further examples and applications of continuous and discrete symmetries can be found in \cite{Leurer:1992wg, Dine:1993np, Nir:1993mx, Leurer:1993gy, Pomarol:1995xc, Barbieri:1998qs, Berezhiani:2000cg, Aranda:2000tm, Lavignac:2001vp, Roberts:2001zy, Ross:2002fb, Nir:2002ah, Kane:2005va, Hirsch:2012ym} (General aspects), \cite{Froggatt:1978nt, Carone:1995xw, Nir:1996am, Binetruy:1996xk, Dudas:1996fe, Barbieri:1996ww, Carone:1997qg, Barbieri:1997tu, Barbieri:1998em, Hall:1998cu, Aranda:1999kc, Barbieri:1999pe, Aranda:2001rd, Chen:2000fp, King:2001uz, Chkareuli:2001dq, Dreiner:2003yr, King:2003rf, Ross:2004qn} (Continuous symmetries), \cite{Kaplan:1993ej, Ma:2001dn, Babu:2002dz, Grimus:2004rj, Grimus:2005mu, Altarelli:2005yx, Hagedorn:2006ug, King:2006np, Feruglio:2007uu, Kubo:2003iw, Kubo:2004ps, Chen:2004rr, Lavoura:2005kx, Teshima:2005bk, Koide:2005ep, Mohapatra:2006pu, Morisi:2005fy, Kaneko:2006wi, Bazzocchi:2007au, Hirsch:2008mg, Hagedorn:2008bc, Feruglio:2009iu, Bazzocchi:2009pv, Bazzocchi:2009da, Hirsch:2009mx, Morisi:2009sc, Hirsch:2010ru, Hagedorn:2010th, Hagedorn:2010mq, Boucenna:2011tj, Toorop:2011jn, deAdelhartToorop:2011ad, Hagedorn:2012pg} (Discrete symmetries).

\section{Charged Lepton Flavor Violation}
\label{sec:lfv}

%
\begin{figure}[t]
\centering
\includegraphics[clip,width=0.80\textwidth]{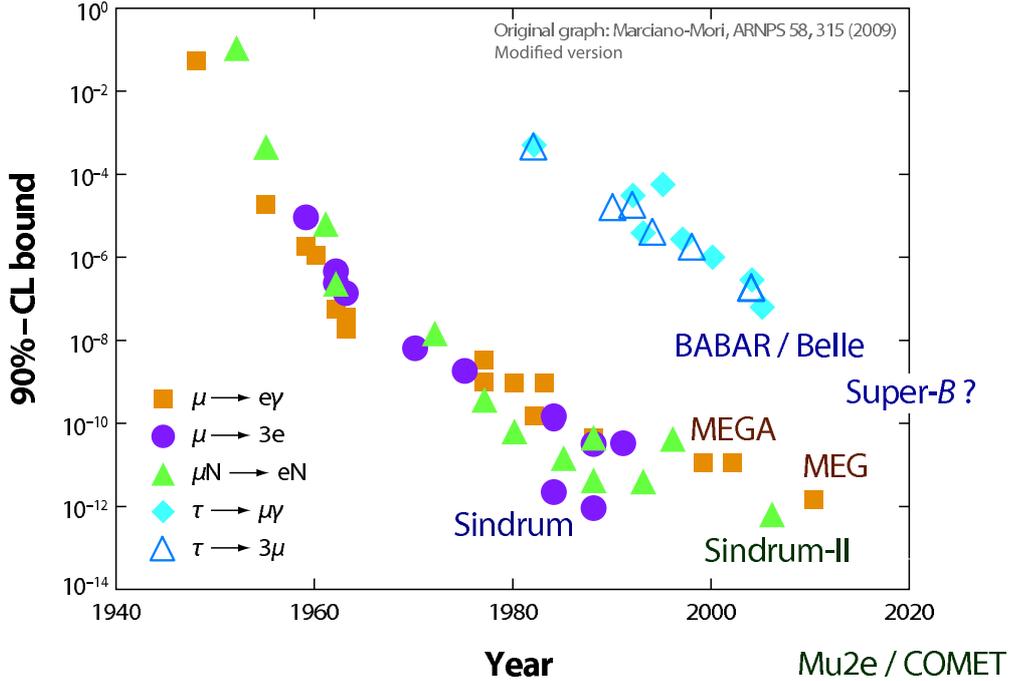}
\caption{History of LFV experiments (from \cite{Hoecker:2012nu}). Displayed are the 90\% C.L. upper limits of past searches as well as the expected sensitivity of future experiments for the indicated LFV processes.}
\label{fig:history}
\end{figure}
Charged lepton flavour violation has been searched for in a large variety of processes such as $\mu\to e\gamma$, $\tau\to e(\mu)\gamma$, $\mu\to eee$ or $\mu-e$ conversion in nuclei. The impressive past and continuing progress of these searches, spanning over 60 years of experimental activity, is displayed in Figure~\ref{fig:history}. So far, no evidence for charged lepton flavour violation has been found. This puts stringent constraints on BSM models which generally contain new sources of LFV, and any successful BSM model must be able to explain how these contributions are suppressed or forbidden. In the following, we will give a brief account of the most stringent current limits and the prospects for future searches. For further information, the reader is referred to the recent overviews of charged lepton flavour physics found in~\cite{Abada:2011rg, Feldmann:2011zh, Hoecker:2012nu}. 
 
\subsection{Lepton Flavour Violating $\mu$ Decays}
\label{sec:lfv_experiment_muegamma}

The MEG experiment~\cite{Adam:2009ci} at PSI (Switzerland) currently provides the most stringent limit on the process $\mu^+\to e^+\gamma$. In this experiment, positive muons with an energy of 29~MeV produced by the $\pi$E5 beam line hit a stopping target at a rate of $3\cdot10^7$~Hz. The MEG detector consists of a positron spectrometer, a positron time-of-flight counter and a scintillation detector, thereby measuring the incidence, decay angles and energies of the photon and the positron. The $\mu^+\to e^+\gamma$ signal consists of back-to-back and mono-energetic ($E=m_{\mu}/2$) pairs of positrons and photons, whereas the background is mainly generated by the accidental coincidence of a positron from the standard decay $\mu^+\to e^+\nu\bar\nu$ with a photon from either a $\mu^+\to e^+\gamma\nu\bar\nu$ decay, bremsstrahlung or positron annihilation. This accidental background increases quadratically with the intensity of the muon beam, which limits this experimental technique in future searches.

Using the data obtained in 2009, a small excess of events in the signal region was reported by the MEG collaboration, which was not reproduced by the 2010 data. Using both data sets, MEG reports a 90\% CL upper limit of~\cite{Adam:2011ch}
\begin{equation}
	\label{eq:MEG}
   Br(\mu^+\to e^+\gamma) 
   \equiv \frac{\Gamma(\mu^+\to e^+\gamma)}{\Gamma(\mu^+\to e^+ \nu\bar\nu)} 
   < 2.4\cdot10^{-12}.
\end{equation}
In a large number of new physics models, this experimental result provides the most stringent constraint on lepton flavour couplings. The current bound (\ref{eq:MEG}) is statistically limited and the MEG experiment is collecting data in 2011 and 2012 to achieve the design sensitivity
\begin{equation}
   Br(\mu^+\to e^+\gamma) \approx \text{few}\times 10^{-13}.
\end{equation}

Similar to $\mu\to e\gamma$, the detection of the decay $\mu\to 3e$ suffers from accidental background from the normal muon decay, but it has the advantage of having only charged particles in the final state and does not require a electromagnetic calorimeter with a more limited performance. The present best limit was reported by the SINDRUM I collaboration in 1987~\cite{Bellgardt:1987du}:
\begin{equation}
	\label{eq:SINDRUM_I}
   Br(\mu^+ \to e^+ e^+ e^-) < 10^{-12}.
\end{equation}
Although, this is numerically lower than the current limit on $\mu\to e\gamma$, Eq.~\ref{eq:MEG}, the latter does provide a more stringent constraint on a large number of new physics models where the LFV contribution from the effective $\mu e\gamma$ vertex dominates, as in this case $\mu\to 3e$ is suppressed by $\approx 10^{-2}$ due to an additional photon vertex. On the other hand, $\mu\to 3e$ can also receive contributions from the direct contact interaction $\mu e e e$. In BSM models where this LFV operator is dominant, such as in R-parity violating SUSY or in Left-Right symmetry with a light doubly charged Higgs, the experimental limit (\ref{eq:SINDRUM_I}) is more constraining. 

In the future, the MuSIC and $\mu3e$ projects could reach a sensitivity of~\cite{Hewett:2012IntensityFrontier}
\begin{equation}
   Br(\mu^+ \to e^+ e^+ e^-) \approx 10^{-16} - 10^{-15}.
\end{equation}

\subsection{$\mu-e$ Conversion in Nuclei}
\label{sec:lfv_experiment_mueconversion}

%
\begin{figure}[t]
\centering
\includegraphics[clip,width=0.70\textwidth]{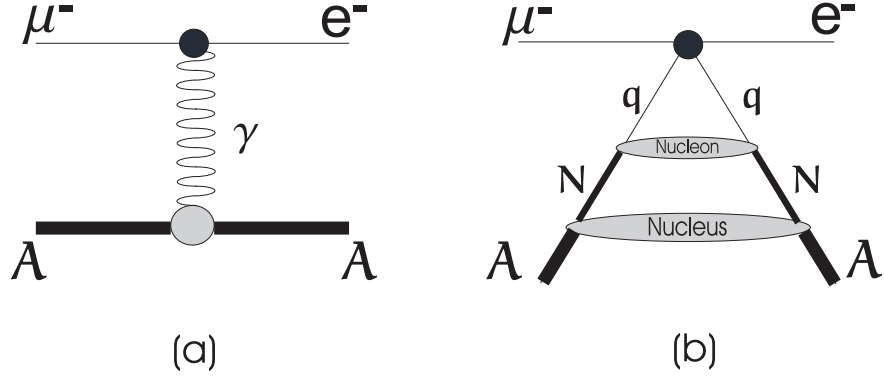}
\caption{Long range (a) and short range (b) contribution to $\mu-e$ conversion (from \cite{Deppisch:2005zm}).}
\label{fig:mue-conv}
\end{figure}
In the presence of LFV operators coupling an electron and muon with a nucleon (such as the electromagnetic coupling of the effective $\mu e\gamma$ dipole vertex with a nucleon, see Fig.~\ref{fig:mue-conv}~(a)), a muon can be captured by a nucleus and undergo a neutrinoless conversion within the nucleus. This process is called $\mu-e$ conversion in a nucleus and to search for it, slow negative muons are guided towards a target where they are captured by an atom and cascade down to 1s orbitals. Controlled by the overlap of the muon wave function with the nucleus, the muon can then convert to an electron, recoiling with the nucleus. Experimentally, the signal simply consists of a single electron with a fixed energy of $\lesssim m_\mu$, with a small, nucleus-dependent deviation from the muon mass due to the muonic atom binding energy and nucleus recoil. Because of the low background of electrons at this energy and because there are no accidentals for such a single particle signal, very high muon rates can be used, giving this technique a potentially high sensitivity. 

The relevant observable measured in the experiment is the ratio of the rate of conversion with the rate of all capturing processes in an atom ${}^A_Z \text{N}$,
\begin{equation}
	R_{\mu e}({}^A_Z \text{N}) = 
	\frac{\Gamma\left(\mu^- + {}^A_Z \text{N} \to e^- + {}^A_Z \text{N} \right)}
	{\Gamma\left(\mu^- + {}^A_Z \text{N} \to {}^A_Z \text{N}^*\right)}.
\end{equation}
Because of the additional electromagnetic coupling of the chirality flipping dipole LFV vertex with the nucleus (Fig.~\ref{fig:mue-conv}~(a)), such a long-range contribution to $\mu-e$ conversion is suppressed by about two orders of magnitude as compared to $\mu\to e\gamma$ (the suppression factor depends on the target nucleus; for Ti, Pb and Al it is given by 1/238, 1/342, 1/389, respectively~\cite{Czarnecki:1998iz,Kitano:2002mt,Marciano:2008zz}). In order to achieve the same sensitivity on new physics couplings and scales, $\mu-e$ conversion experiments must therefore be able to probe smaller conversion rates $R_{\mu e}({}^A_Z N) \approx \text{few }10^{-3}Br(\mu\to e\gamma)$. On the other hand, $\mu-e$ conversion receives contributions from effective contact interactions coupling $\mu e q q$ (Fig.~\ref{fig:mue-conv}~(b)), e.g. arising from box diagrams with heavy particles in the loop (see Fig.~\ref{fig:diagrams_LFV_mue} in Left-Right symmetric models). Such operators do not contribute to $\mu\to e\gamma$~\cite{Raidal:2008jk}, and they enhance the $\mu-e$ conversion rate with respect to $\mu\to e\gamma$. The ratio $Br(\mu\to e\gamma)/R_{\mu e}({}^A_Z N)$ as well as the ratio of $\mu-e$ conversion rates in different nuclei is therefore model-independent and, if these processes are discovered, can be used to distinguish between new physics mechanisms~\cite{Raidal:2008jk,Cirigliano:2009bz}.

The current best limit on $\mu-e$ conversion is given by~\cite{Bertl:2006up}
\begin{equation}
	R_{\mu e}(\text{Au}) < 7\times 10^{-13},
\end{equation}
achieved in 2006 by the SINDRUM--II collaboration at PSI using gold. In the future, the proposed experiments Mu2e (FNAL)~\cite{Kutschke:2011ux} and COMET (J-PARC)~\cite{Kurup:2011zza} both aim to reach a sensitivity of 
\begin{equation}
	R_{\mu e}(\text{Al}) \approx 10^{-16}
\end{equation}
by the end of the decade. This exceeds the equivalent sensitivity of the expected MEG reach of  $Br(\mu^+\to e^+\gamma) \approx 10^{-13}$ for dipole contribution dominated new physics scenarios, and would make these experiments the most sensitive probes of lepton flavour violation. In the longer term, these projects could be upgraded to improve the sensitivity even further by two orders of magnitude, utilising the Project-X proton accelerator at FNAL~\cite{Tschirhart:2011zza} or the PRISM/PRIME project~\cite{Barlow:2011zz}, respectively.

\subsection{Lepton Flavour Violating $\tau$ Decays}
\label{sec:lfv_experiment_taudecays}

Searching for flavour violating transitions $\tau\to e$ and $\tau\to\mu$ is experimentally much more difficult due to the lower fluxes in $\tau$ production which are available so far. This experimental fact is especially problematic for the power of theoretical frameworks which aim to predict the relative strengths of these transitions, such as for flavour symmetries as discussed in this contribution.

\begin{figure}[t]
\centering
\includegraphics[clip,width=0.90\textwidth]{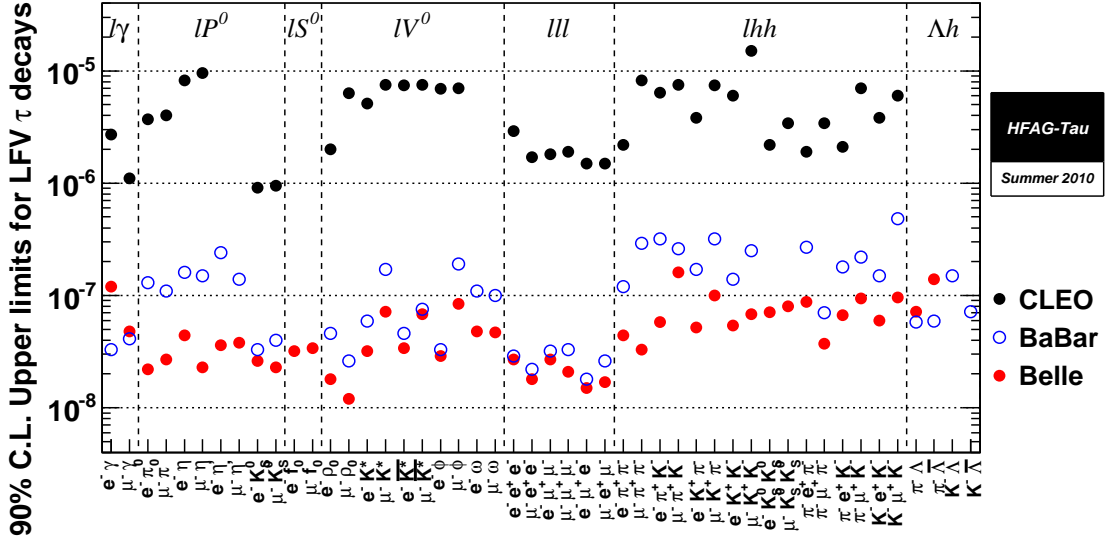}
\caption{Upper limits on the branching ratios of lepton flavor violating $\tau$ decays reported by the BABAR, Belle and CLEO collaborations (from \cite{Asner:2010qj}).}
\label{fig:TauLFVLimits}
\end{figure}
The current best limits on flavour violating $\tau$ decays are obtained by the B-factories BABAR and Belle, which in total searched for 48 different LFV final states~\cite{Asner:2010qj}, see Fig.~\ref{fig:TauLFVLimits}. Generally most relevant for phenomenology are the decays to $\ell\gamma$ with the 90\% C.L. limits (charge averaged)~\cite{Aubert:2009ag}
\begin{align}
	Br(\tau\to e\gamma)   &< 3.3\times 10^{-8},\\
	Br(\tau\to \mu\gamma) &< 4.4\times 10^{-8},
\end{align}
and the decays to three-lepton final states ranging between~\cite{Miyazaki:2011xe}
\begin{equation}
	Br(\tau^-\to \ell_1^- \ell_2^+ \ell_3^-) \lesssim (1.5 - 3.0)\times 10^{-8}, \quad
	\ell_1,\ell_2,\ell_3 = e,\mu,
\end{equation}
depending on the composition of the final state (cf.~Fig.~\ref{fig:TauLFVLimits}). In the future, the Super B-factory projects SuperB~\cite{Bona:2007qt} and Belle II~\cite{Abe:2010sj} could achieve sensitivities on LFV $\tau$ decays of the order~\cite{1742-6596-171-1-012079}
\begin{equation}
	Br_\text{Super B}(\tau\to \ell\gamma) \sim \text{few}\times 10^{-9}, \quad \ell=e,\mu.
\end{equation}
Taus are also copiously produced at the LHC, and searches for the decay $\tau\to\mu\mu\mu$ could be competitive to current limits~\cite{Giffels:2008ar}. At LHCb, taus are dominantly produced in the decays of $B$ and $D$ mesons. Using data collected in 2011 with $\sqrt{s} = 7$~TeV and the luminosity $\mathcal{L} = 1.0\text{ fb}^{-1}$, the LHCb collaboration has reported an upper limit $Br(\tau^- \to \mu^- \mu^- \mu^+) < 6.3\times 10^{-8}$ (90 \% C.L.)~\cite{Seyfert:1451298, LHCb-CONF-2012-015}. This is already starting to be competitive with the sensitiviy achieved at the B-factories.

\subsection{Lepton Flavour Violating Collider Processes}
\label{sec:lfv_experiment_collider}

As noted above, lepton flavour violating $\tau$ decays can be searched for at the LHC, although it is not expected that the sensitivity would improve compared to dedicated experiments at the luminosity frontier, such as B-factories. On the other hand, LFV signatures at the LHC can arise in the production and decay of new heavy particles in BSM models. As the decay products of such heavy particle are generally highly energetic, the reconstruction of $\tau$'s produced in such a way, although challenging in its own right, is simplified. The experimental acceptance of LFV signatures $\tau\ell + X$ $(\ell=e,\mu)$, can then be $\approx 25$\% \cite{Giffels:2008ar,AguilarSaavedra:2012fu}. Effectively, the sensitivity to $\tau\ell + X$ LFV signatures would only be reduced by a factor of $\approx 4$ compared to $\mu e + X$. This is a much more favourable ratio than for low-energy observables, $Br(\tau\to\ell\gamma) : Br(\mu\to e\gamma)  \gtrsim 10^3$. In general, searches for LFV signatures at the LHC can be competitive to low-energy experiments as long as the mediating new physics particles can be produced, although this is highly model- and parameter-dependent. Examples fo such scenarios will be discussed in more detail below.

\section{New Physics Models}
\label{sec:models}

\subsection{Effective Operator Approach}
\label{sec:models_effective}

Before discussing specific realisations of BSM theories, we will briefly introduce a highly useful approach to understand and categorise new physics effects on low-energy observations. In an effective low-energy description, such effects are parametrised by expanding the SM Lagrangian with higher-dimensional operators. If new physics enters at a scale $M_\text{NP}$, an effective Lagrangian can be be written as as an expansion in inverse powers of $M_\text{NP}$,
\begin{equation}
\label{eq:Leff}
	\mathcal{L}_\text{eff} = 
	\mathcal{L}_\text{SM} + 
	\frac{1}{M_\text{NP}}   \mathcal{L}^5 +
	\frac{1}{M_\text{NP}^2} \mathcal{L}^6 +  
	\dots,
\end{equation}
where $\mathcal{L}_\text{SM}$ is the renormalisable SM Lagrangian, and the terms $\mathcal{L}^d$ contain all possible non-renormalizable effective operators, $\mathcal{L}^d = \sum_n C^d_n\mathcal{O}^d_n(\text{SM fields}) + \text{h.c.}$, of mass dimension $d$. Here, $C^d_n$ are coefficients which can be calculated within a specific new physics model at $M_\text{NP}$ generating the respective operator by integrating out the heavy particles. The effective operators must be invariant under the SM gauge group $SU(3) \times SU(2) \times U(1)$, and are written as combinations of SM fields. In the following, we will concentrate on dimension 5 and 6 operators contributing to LFV processes. Higher dimensional operators of dimension 7 and above can provide interesting effects such as neutrino transition moments and Higgs-induced LFV \cite{Raidal:2008jk}. Although suppressed by additional powers of $M_\text{NP}$ these operators can be relevant if new physics enters close to scales accessible at the LHC.

At dimension 5, the only operator allowed (modulo flavour indices) is the famous Weinberg operator, 
\begin{equation}
	\label{eq:WeinbergOperator}
	\mathcal{O}^5_{ij} = \frac{1}{2}\kappa^\nu_{ij} 
	(\bar L_i\cdot H) (H^\dagger\cdot L_j)^c + \text{h.c.},
\end{equation}
generating light Majorana neutrino masses after electroweak symmetry breaking, $m^\nu_{ij} = \kappa^\nu_{ij} \langle H_0\rangle$. As indicated in the introduction, this operator does not induce sizable LFV due to the extreme lightness of the observed neutrino masses. 

At dimension 6, the operators relevant for LFV processes can be categorized and schematically written as 
\begin{enumerate}
\item Two-Lepton - Higgs - Photon operator:
\begin{equation} 
	{}^{e\gamma}\mathcal{O}^{6}_{ij}  = 
	\bar L_i \sigma^{\mu\nu} e^c_j H^\dagger F_{\mu\nu}
\end{equation}
After the Higgs field acquires its VEV, these operators give rise to chirality flipping dipole moments of the charged leptons (transitional moments if $i\neq j$). Specifically, the flavour diagonal component ${}^{e\gamma}\mathcal{O}^{6}_{ii}$ at zero-momentum transfer yields anomalous magnetic and electric dipole moments of lepton $i$, proportional to the real and imaginary part of the corresponding expansion coefficient ${}^{e\gamma}\mathcal{C}^{6}_{ij}/M_{NP}^2$, respectively. The flavour off-diagonal components at zero-momentum transfer ${}^{e\gamma}\mathcal{O}^{6}_{ij,ji}$ give rise to the LFV decay $l_i\to l_j\gamma$. 
\item Four-Lepton operators of the form
\begin{equation}
	{}^{4e}\mathcal{O}^{6}_{ijkl}  = 
	(\bar e_i \gamma^\mu P_1 e_j)(\bar e_k \gamma_\mu P_2 e_l), \quad
	(P_1, P_2) = (P_R, P_R), (P_L, P_L), (P_R, P_L).
\end{equation}
Most relevant to our discussion, the flavour violating components of these operators give rise to LFV decays such as $\mu\to 3e$, $\tau\to 3\mu$, etc. depending on the flavour combination $ijkl$.

\item Two-Lepton - Two-Quark operators of the form
\begin{equation}
	{}^{eq}\mathcal{O}^{6}_{ijkl}  = 
	(\bar e_i \gamma^\mu P_L e_j)(\bar q_k \gamma_\mu P_L q_l), \quad
	(\bar e_i \gamma^\mu P_R e_j)(\bar q_k \gamma_\mu P_R q_l), \quad
	(\bar e_i P_R e_j)(\bar q_k P_L q_l).
\end{equation}
These operators can give contributions to a host of different processes, such as $\mu-e$ conversion in nuclei, LFV meson decays and corrections to quark flavor observables.
\end{enumerate}
As indicated above, these operators contribute to a range of low-energy physics processes, and the corresponding effective coupling coefficients $C^6_n/M_{NP}^2$ are constrained by experimental data, in many cases stringently. For example, the branching ratio $Br(l_i\to l_j\gamma)$ can be expressed in terms of the coefficients $C_{ij}$ of the dipole operators ${}^{e\gamma}\mathcal{O}^6_{ij}$ as~\cite{Altarelli:2010gt}
\begin{equation}
	\label{eq:BrMuEGammaEff}
	Br(l_i\to l_j\gamma) = \frac{12\sqrt{2}\pi^3\alpha}{G_F^3 m_{l_i}^2 M_\text{NP}^2}
	\left(
		\left|C_{ij}\right|^2 +
		\left|C_{ji}\right|^2
	\right),
\end{equation} 
and the current limits on the LFV decays $l_i\to l_j\gamma$ correspond to
\begin{align}
	Br(\mu\to e\gamma) < 2.4 \times 10^{-12} &\Rightarrow
	\left|C_{\mu e}\right| \lesssim 5\times 10^{-9} 
		\times \left(\frac{1\text{ TeV}}{M_\text{NP}}\right)^2, \nonumber\\
	Br(\tau\to \ell\gamma) \lesssim 4.0 \times 10^{-8\phantom{1}} &\Rightarrow
	\left|C_{\tau\ell}\right| \lesssim 6 \times 10^{-7} 
		\times \left(\frac{1\text{ TeV}}{M_\text{NP}}\right)^2, \quad\ell = e,\mu.
\end{align} 
The effective couplings are therefore severely constrained if new physics responsible for mediating these processes would be present close to the TeV scale, which would hint at a suppression mechanism such as a flavour symmetry. Vice versa, if we assume coefficients of order one, the scale of new physics has to be many orders of magnitude above the electroweak scale, $M_\text{NP} \gtrsim 10^4$~TeV. Similar conclusions can be drawn for the other operators and the corresponding observables. An exhaustive list is provided in \cite{Raidal:2008jk} and the recent analysis \cite{Calibbi:2012at}. Although the effective operator approach is highly general in describing new physics effects at low-energies, one has to be careful in interpreting the limits with respect to specific models~\cite{Raidal:2008jk}: (i) Usually, the limit on a coefficient is calculated assuming the dominance of only one operator, but new physics generally results in many effective operators which can interfere and cancel. (ii) As mentioned before, contributions from higher-dimensional operators might be neglected which could be sizeable if there is new physics around the TeV scale. (iii) There may be more than one new physics scale generating different operators. For example, in Seesaw scenarios where the Weinberg operator (\ref{eq:WeinbergOperator}) is generated through the breaking of lepton number close to GUT scale $M_\text{LNV} \lesssim 10^{15}$~GeV, while LFV effects are generated at a much lower scale $M_\text{LFV} \gtrsim 1$~TeV (such as the SUSY mass scale), the effective Lagrangian (\ref{eq:Leff}) should be expanded as~\cite{Feldmann:2011zh}
\begin{equation}
\label{eq:Leff2}
	\mathcal{L}_\text{eff} = 
	\mathcal{L}_\text{SM} + 
	\frac{1}{M_\text{LNV}}   \mathcal{L}^5 +
	\frac{1}{M_\text{LFV}^2} \mathcal{L}^6 +  
	\dots.
\end{equation}

Applying the effective operator approach to flavour symmetries, we consider the scenario~\cite{Altarelli:2010gt} where a new physics model is invariant under a flavour symmetry above the scale $M_\text{F}$. When breaking the flavour symmetry through the VEV(s) $v_i$ of flavon(s), effective LFV operators are generated with coefficients $C^6_n$ that depend on the small parameters $\epsilon_i = v_i/M_{F}$. By establishing the dependence of the operator coefficients on these breaking parameters, it is possible to predict the relative strength of operators with different flavour transitions. This technique can be used to distinguish between different flavour symmetries without explicitly fixing the new physics model. We will highlight the use of this method for two examples presented in~\cite{Altarelli:2010gt, Merlo:2010mw}.

In the model with flavour symmetry $A_4 \times Z_3 \times U(1)_\text{FN}$ of Section~\ref{sec:A4}, the coefficients $C_{ij}$ of the dipole operators ${}^{e\gamma}\mathcal{O}^{6}_{ij}$ are given by~\cite{Feruglio:2009hu, Hagedorn:2009df}
\begin{equation}
	\label{eq:CA4}
	C \approx 
	\begin{pmatrix}
		\lambda^2 \epsilon   & \lambda^2 \epsilon^2 & \lambda^2 \epsilon^2 \\
		\lambda   \epsilon^2 & \lambda   \epsilon   & \lambda   \epsilon^2 \\
		          \epsilon^2 &           \epsilon^2 &           \epsilon
	\end{pmatrix},
\end{equation}
with the $U(1)_\text{FN}$ breaking parameter $\lambda = \langle \theta_\text{FN} \rangle / M_F$, the $A_4$ breaking parameter $\epsilon = \langle\varphi_T\rangle / M_F$ and every entry is assumed to contain coefficients $c_{ij} = \mathcal{O}(1)$. To lowest order, in the small parameters $\lambda$ and $\epsilon$, the branching ratio (\ref{eq:BrMuEGammaEff}) always scales as $\epsilon^2/M_\text{NP}^4$. The three LFV decay rates are therefore predicted to be of the same order in this model:
\begin{equation}
	Br(\mu\to e\gamma) \approx Br(\tau\to e\gamma) \approx Br(\tau\to\mu\gamma).
\end{equation}

In a model with flavour symmetry $S_4 \times Z_3 \times U(1)_\text{FN}$, the dipole operator coefficient matrix is given by~\cite{Altarelli:2010gt} (up to factors of $\mathcal{O}(1)$)
\begin{equation}
	\label{eq:CS4}
	C \approx 
	\begin{pmatrix}
		\lambda^2 \epsilon^2             & 
		\lambda^2 \epsilon^2 \epsilon'   & 
		\lambda^2 \epsilon^2 \epsilon    \\
		\lambda   \epsilon   \epsilon'   & 
		\lambda   \epsilon               & 
		\lambda   \epsilon   \epsilon'^2 \\
		          \epsilon   \epsilon'   &           
		          \epsilon   \epsilon'^2 &
		          \epsilon
	\end{pmatrix},
\end{equation}
where $\lambda = \langle \theta_\text{FN} \rangle / M_F \ll 1$ is again the $U(1)_\text{FN}$ breaking parameter whereas $\epsilon, \epsilon' \ll 1$ are associated with the breaking of $S_4$. In this model, $Br(\mu\to e\gamma)$ and $Br(\tau\to e\gamma)$ scale as $(\epsilon\epsilon')^2/M_\text{NP}^4$ to lowest order in the small parameters while $Br(\tau\to\mu\gamma)$ scales as $(\epsilon^2\epsilon'^4)/M_\text{NP}^4$, leading to the prediction
\begin{equation}
	Br(\mu\to e\gamma) \approx Br(\tau\to e\gamma) \gg Br(\tau\to\mu\gamma).
\end{equation}

\subsection{SUSY Seesaw}
\label{sec:models_susy}

One of the most popular BSM frameworks to study LFV is the supersymmetric seesaw mechanism. The most elegant explanation for small neutrino masses is provided by the see-saw mechanism~\cite{gell-mann:1980vs, Minkowski:1977sc, Glashow:1979, Mohapatra:1979ia, Yanagida:1979, Schechter:1980gr, Mohapatra:1981yp}, in which a large Majorana mass scale $M_R$ of right-handed neutrinos drives the light neutrino masses down to or below the sub-eV scale, as required by the experimental evidence. A priori, the fundamental scale $M_R$ can be of the order of the GUT scale, and may thus be unaccessible for any kind of direct experimental tests. However, neutrino mixing implies lepton flavour violation (LFV), which is absent in the Standard Model and provides indirect probes of $M_R$. While lepton flavour violating processes are suppressed due to the small neutrino masses if only right-handed neutrinos are added to the Standard Model, in supersymmetric models new sources of LFV exist. For example, virtual effects of the massive neutrinos affect the renormalization group equations (RGE) of the slepton mass and the trilinear coupling matrices, and give rise to non-diagonal terms inducing LFV. One finds that flavour violation in the neutrino Yukawa couplings is indeed transmitted to charged leptons through charged slepton and sneutrino loops, giving a sizeable enhancement to LFV process rates~\cite{Hall:1986dx, Borzumati:1986qx, Barbieri:1994pv}. LFV processes in the context of supersymmetric seesaw models have been considered in several previous studies (see e.g. \cite{Hisano:1996cp, Hisano:1999fj, Casas:2001sr, Kageyama:2001tn, Deppisch:2002vz, Deppisch:2002tv, Deppisch:2002qw} and references therein). In \cite{Hisano:1999fj} it has been pointed out that the corresponding branching ratios and cross-sections exhibit a quadratic dependence on the right-handed Majorana neutrino mass scale $M_{R}$.

\begin{figure}[t]
\centering
\begin{picture}(400,100)(-190,-25)
\ArrowLine(-190,0)(-150,0)
\Text(-170,-10)[]{$l_{i}$}
\DashLine(-150,0)(-80,0){5}
\Text(-135,-10)[]{$\tilde{l}$}
\Photon(-115,0)(-115,-45){3}{5}
\Text(-100,-30)[]{$\gamma$}
\ArrowLine(-80,0)(-40,0)
\Text(-60,-10)[]{$l_{j}$}
\CArc(-115,0)(35,0,180)
\Text(-115,46)[]{$\tilde{\chi}^{0}$}
\ArrowLine(-20,0)(20,0)
\DashLine(20,0)(90,0){5}
\ArrowLine(90,0)(130,0)
\Text(00,-10)[]{$l_{i}$}
\Text(55,-10)[]{$\tilde{\nu}$}
\Photon(55,35)(55,74){3}{5}
\Text(70,55)[]{$\gamma$}
\Text(110,-10)[]{$l_{j}$}
\CArc(55,0)(35,-360,-180)
\Text(28,34)[]{$\tilde{\chi}^{-}$}
\end{picture} 
\vspace*{1cm}
\caption{\label{lfv_lowenergydiagrams} Diagrams for $l_{i}^{-}\rightarrow l_{j}^{-}\gamma$ in the MSSM (from \cite{Deppisch:2002qw}).}
\end{figure}
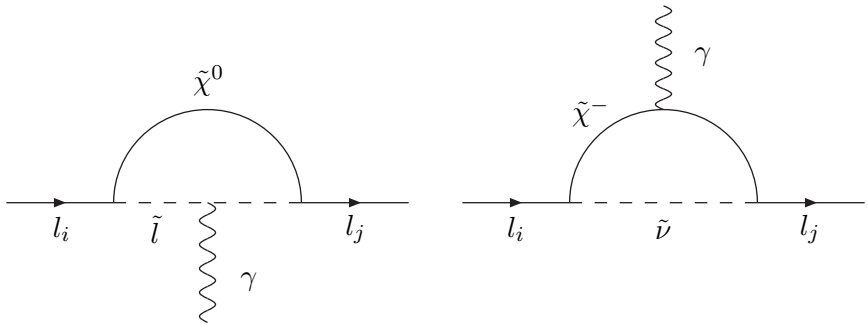
If three right-handed neutrino singlet fields $\nu_R$ are added to the particle content of the Minimal Supersymmetric Standard Model (MSSM), one gets the following additional terms in the superpotential:
\begin{equation}
\label{suppot4}
	W_\nu = -\frac{1}{2}\nu_R^{cT} M \nu_R^c + \nu_R^{cT} Y_\nu L \cdot H_u.
\end{equation}
Here, \(Y_\nu\) is the matrix of neutrino Yukawa couplings, $M$ is the right-handed neutrino Majorana mass matrix, and $L$ and $H_u$ denote the left-handed lepton and hypercharge +1/2 Higgs doublets, respectively. At energies much below the mass scale of the right-handed neutrinos, $W_{\nu}$ leads to the mass matrix 
\begin{equation}
\label{eqn:SeeSawFormula}
	M_\nu = m_D^T M^{-1} m_D = Y_\nu^T M^{-1} Y_\nu (v \sin\beta )^2,
\end{equation}
for the left-handed neutrinos. Thus, light neutrino masses are naturally obtained if the typical scale of the Majorana mass matrix \(M\) is much larger than the scale of the Dirac mass matrix \(m_D=Y_\nu \langle H_u^0 \rangle\),  where \(\langle H_u^0 \rangle = v\sin\beta\) is the appropriate Higgs VEV with \(v=174\)~GeV and \(\tan\beta =\frac{\langle H_u^0\rangle}{\langle H_d^0\rangle}\).  The dominantly right-handed neutrino mass eigenstates are too heavy to be observed directly. However, they give rise to virtual corrections to the slepton mass matrices that can be responsible for observable lepton-flavour violating effects. In particular, the soft SUSY mass term $m_L^2$ of the left-handed slepton doublet and the trilinear charged slepton couplings $A$ are affected by the presence of neutrino Yukawa couplings above the seesaw scale. Renormalising the slepton mass terms from the GUT scale $M_\text{GUT}$ to the electroweak scale one obtains with mSUGRA boundary conditions at $M_\text{GUT}$
\begin{eqnarray}
	m_{L}^2 &=& m_0^2\mathbf{1} 
	+ (\delta m_{L}^2)_{\textrm{\tiny MSSM}} + \delta m_{L}^2
	\label{left_handed_SSB} \nonumber\\
	A &=& A_0 Y_l+\delta A_{\textrm{\tiny MSSM}}+\delta A \label{A_SSB},
\end{eqnarray}
where $m_{0}$ is the common soft SUSY-breaking scalar mass and $A_{0}$ the common trilinear coupling. The terms \((\delta m_L^2)_{\textrm{\tiny MSSM}}\) and \(\delta A_{\textrm{\tiny MSSM}}\) are flavour-diagonal corrections already present in the MSSM. In addition, the evolution generates off-diagonal terms in $\delta m_L^2$ and $\delta A^2$  which in leading-log approximation are given by
\begin{eqnarray}
\label{eq:rnrges}
	\delta m_{L}^2 &=& -\frac{1}{8 \pi^2}(3m_0^2+A_0^2)(Y_\nu^\dagger L Y_\nu), 
	\quad\quad L_{ij}=\ln\left(\frac{M_\text{GUT}}{M_{i}}\right)\delta_{ij},
	\label{left_handed_SSB2}   \nonumber\\
	\delta A &=&  -\frac{3 A_0}{16\pi^2}(Y_l Y_\nu^\dagger L Y_\nu),
\end{eqnarray}
where $M_i,~i=1,2,3$ are the eigenvalues of the Majorana mass matrix $M$.

The product of the neutrino Yukawa couplings $Y_\nu^\dagger L Y_\nu$ entering these corrections can be written in terms of the light neutrino masses $m_{\nu_i}$ and mixing matrix $U$ as
\begin{eqnarray}
\label{eqn:yy}
	Y_\nu=\frac{1}{v\sin\beta}\textrm{diag}
	\left(\sqrt{M_1}, \sqrt{M_2}, \sqrt{M_3}\right) R \; 
	\textrm{diag}
	\left(\sqrt{m_{\nu_1}}, \sqrt{m_{\nu_2}}, 
	\sqrt{m_{\nu_3}}\right)U_\text{PMNS}^\dagger, 
\label{eq:yukawa}
\end{eqnarray}
where $R$ is an unknown complex orthogonal matrix~\cite{Casas:2001sr}. For real $R$ and degenerate right-handed Majorana masses, $R$ as well as the light neutrino Majorana phases drop out from the product $Y_\nu^\dagger L Y_\nu$. In this scenario, the Yukawa term is given by~\cite{Deppisch:2002vz}
\begin{eqnarray}
	\left(Y_{\nu}^{\dagger}L Y_{\nu}\right)_{ab}
	& \approx & \frac{M_{R}}{v^{2}\sin^{2}\beta}
	\left(\sqrt{\Delta m^{2}_{12}} U_{a2}U_{b2}^{*} 
	+ \sqrt{\Delta m_{23}^2}U_{a3}U_{b3}^{*}\right)
	\ln\frac{M_\text{GUT}}{M_{R}}.
\label{llcorrectionhier} 
\end{eqnarray}
This result can also be understood as the presence of a flavour symmetry, namely the assumption of minimal flavour violation (MFV)~\cite{Chivukula:1987py, Hall:1990ac, D'Ambrosio:2002ex}. In this framework, the flavour coefficients of the slepton mass matrix can be expressed in terms of the charged lepton Yukawa couplings and the neutrino mass matrix by requiring that the sources of LFV in new physics (in this case, the charged slepton and sneutrino mass matrices) are invariant under the minimal flavour symmetry $U(3)_{e^c} \times U(3)_L$. This symmetry is only to be broken by the charged lepton Yukawa couplings and the couplings of the Weinberg operator generating the light neutrino masses. For more details on this analogy, see \cite{Cirigliano:2005ck, Raidal:2008jk} and references therein.

At low energies, the flavour off-diagonal correction (\ref{left_handed_SSB2}) induces
the radiative decays \(l_i\to l_j \gamma\), via the photon penguin diagrams shown in Fig.~\ref{lfv_lowenergydiagrams}, with charginos/sneutrinos or neutralinos/charged sleptons in the loop \footnote{In our discussion, we focus on the LFV decays $l_i \to l_j \gamma$, induced by a dominant photon dipole operator. The reader should note, though, that the corresponding $Z$ penguin is not negligible in many extensions of the MSSM \cite{Hirsch:2012ax, Abada:2012cq}. This can lead to enhanced rates of processes such as $\mu \to 3 e$ and $\mu-e$ conversion in nuclei compared to $\mu \to e \gamma$.}. Under the assumptions leading to (\ref{llcorrectionhier}), one always has $Br(\mu\to e\gamma) < Br(\tau\to\mu\gamma)$. The possibility of cancellations of contributions in (\ref{llcorrectionhier}) to $Br(\mu\to e\gamma)$ for small values of $\theta_{13}$ is disfavoured in light of the recent evidence for $\sin^2 2\theta_{13} \approx 0.1$~\cite{An:2012eh, Ahn:2012nd}.

\begin{figure}[t]
\begin{center}
\setlength{\unitlength}{1cm}
\includegraphics[clip,width=0.5\textwidth]{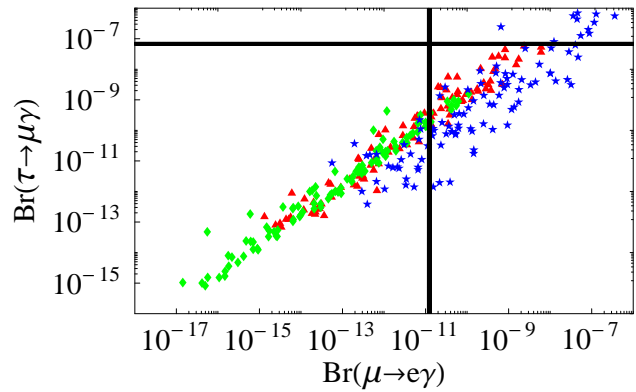}
\end{center}
\caption{$Br(\tau\to\mu\gamma)$ versus $Br(\mu\to e\gamma)$ in mSUGRA scenario SPS1a with neutrino parameters scattered within their experimentally allowed ranges (from \cite{Raidal:2008jk}). For quasi-degenerate heavy neutrino masses, both hierarchical (triangles) and quasi-degenerate (diamonds) light neutrino masses are considered with real $R$ and $10^{11}\text{ GeV} < M_R < 10^{14.5}$~GeV. In the case of hierarchical heavy and light neutrino masses (stars), the parameters of the matrix $R$ are scattered over their full ranges demanded by leptogenesis and perturbativity~\cite{Deppisch:2005rv}.}
\label{fig:Br12Br23}
\end{figure}
In Figure~\ref{fig:Br12Br23}, the correlation between $Br(\tau\to\mu\gamma)$ and  $Br(\mu\to e\gamma)$ is displayed for a specific minimal supergravity inspired MSSM (mSUGRA) scenario with three different choices for the light and heavy neutrino mass spectra as well as the matrix $R$. In all cases, $Br(\mu\to e\gamma)$ is a much more sensitive probe, and an observation of $Br(\tau\to\mu\gamma)$ in the foreseeable future is ruled out in this framework. The updated current limit on $Br(\mu\to e\gamma) < 2.4\times 10^{-12}$ only strengthens this observation. Similar results hold for the decay $\tau\to e\gamma$ which is also severely constrained by the non-observation of $\mu\to e\gamma$. This is also the case in many SUSY seesaw models with specific flavour symmetries. As demonstrated in Section~\ref{sec:models_effective} for the case of two example flavour groups $A_4 \times Z_3 \times U(1)_\text{FN}$ and $S_4 \times Z_3 \times U(1)_\text{FN}$, $Br(\mu\to e\gamma)$ is equal to or larger than the other LFV decays. Examples of analyses of these and similar flavour symmetries (also taking into account higher order corrections) in the SUSY seesaw framework, which support this general observation, can be found in~\cite{Ding:2009gh, Feruglio:2009hu, Hagedorn:2009df, Ishimori:2010su}. Nevertheless, in specific choices of flavour symmetries, the rate of $\mu\to e\gamma$ can vary wildly for a given SUSY parameter point. For example, in a supersymmetric $A_4$ flavour model analyzed in \cite{Hirsch:2003dr}, large $\tau-\mu$ LFV rates are predicted with $Br(\tau\to\mu\gamma) > 10^{-9}$ for $Br(\mu\to e\gamma) < 10^{-11}$. If the absolute SUSY scale is known, such as by discovery at the LHC, this could be used to distinguish between flavour symmetries. This was for example demonstrated in \cite{Deppisch:2010sv} in a  scan over Froggatt-Nielsen inspired flavour textures with tri-bimaximal neutrino mixing.

At the LHC, a feasible test of LFV is provided by production of squarks and gluinos, followed by cascade decays via neutralinos and sleptons. LFV can occur in the decay of the second lightest neutralino and the slepton, resulting in different lepton flavours. The total cross section for the signature \(l^+_i l^-_j + X\) can then be written as
\begin{align}
\label{eqn:LHCProcess}
   &\sigma(pp\to l^+_i l^-_j+X) =                             \nonumber\\
   &\Bigl\{
   	\sum_{a,b}\sigma(pp\to\tilde q_a\tilde q_b)\times 
   	Br(\tilde q_a\to\tilde\chi^0_2 q_a)
   	+ \sum_{a}\sigma(pp\to\tilde q_a\tilde g)  \times 
   	Br(\tilde q_a\to\tilde\chi^0_2 q_a)                     \nonumber\\
   	&+ \sigma(pp\to\tilde g\tilde g)\times 
   	Br(\tilde g\to\tilde\chi^0_2 g) 
   	\Bigr\}
   	\times Br(\tilde\chi^0_2\to l_i^+l_j^-\tilde\chi^0_1),
\end{align}
where \(X\) can involve jets, leptons and LSPs produced by lepton flavour conserving decays of squarks and glui\-nos, as well as low energy proton remnants.

\begin{figure}[t]
\centering
\includegraphics[clip,width=0.49\textwidth]{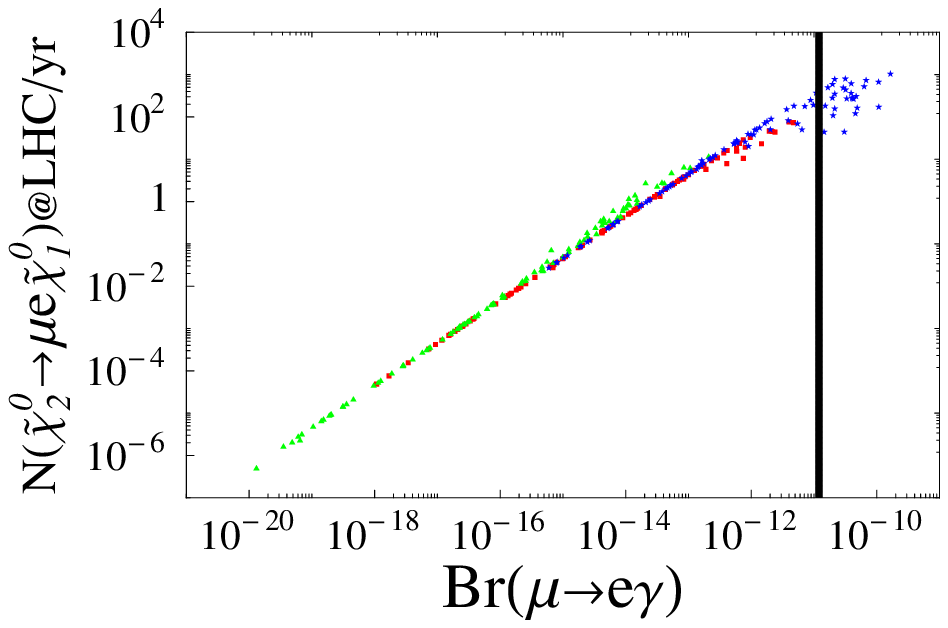}
\includegraphics[clip,width=0.49\textwidth]{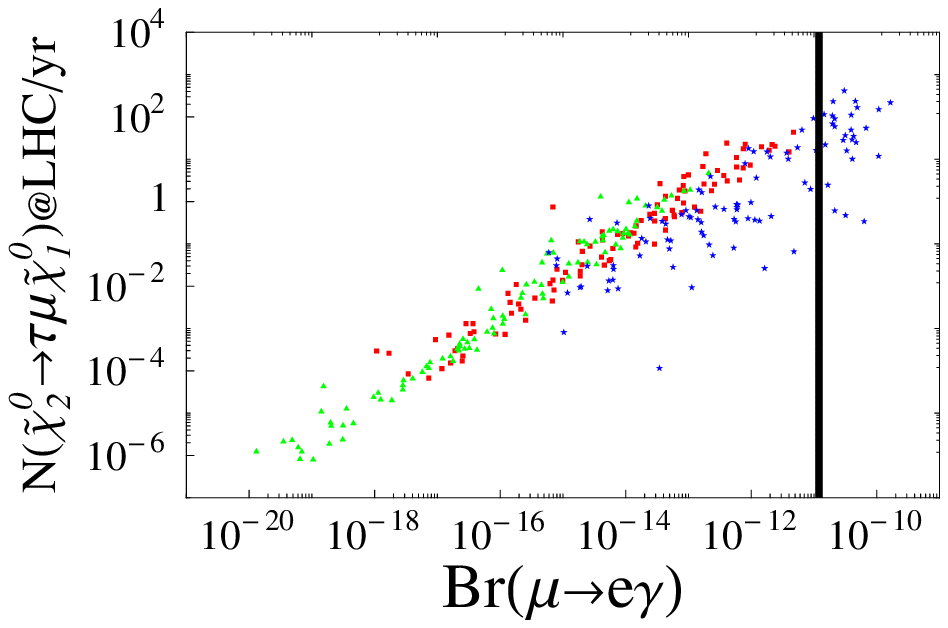}
\caption{Correlation of the number of \(\tilde\chi_2^0\to\mu^+e^-\tilde\chi_1^0\) (left) and \(\tilde\chi_2^0\to\tau^+\mu^-\tilde\chi_1^0\) events at the LHC with 14~TeV and \(100\textrm{ fb}^{-1}\) with \(Br(\mu\to e\gamma)\) in the mSUGRA scenario (\(m_0=85\)~GeV, \(m_{1/2}=400\)~GeV, \(A_0=0\)~GeV, \(\tan\beta=10\)~GeV, \(\textrm{sign}\mu=+\)) for the case of hier.\ $\nu_{R/L}$ (blue stars), deg.\ $\nu_R$/hier.\ $\nu_L$ (red
boxes) and deg.\ $\nu_{R/L}$ (green triangles) (from \cite{Raidal:2008jk}). The neutrino parameters are scattered within their experimentally allowed ranges~\cite{Maltoni:2004ei}. For degenerate heavy neutrino masses, both hierarchical (green diamonds) and degenerate (blue
stars) light neutrino masses are considered with real $R$ and $10^{11}\ {\rm GeV}<M_R < 10^{14.5}\ {\rm GeV}$. In the case of hierarchical heavy and light neutrino masses (red triangles), the parameters of the $R$ matrix are scattered in the ranges allowed by leptogenesis and perturbativity~\cite{Deppisch:2005rv}.}
\label{fig:br12_N2}
\end{figure}
As a sensitivity comparison it is useful to correlate the expected LFV event rates at the LHC with LFV rare decays. This is shown in Fig.~\ref{fig:br12_N2} for the event rates \(N(\tilde\chi_2^0 \to \mu^+e^-\tilde\chi_1^0)\) and \(N(\tilde\chi_2^0 \to \tau^+\mu^-\tilde\chi_1^0)\). Both are correlated with \(Br(\mu\to e\gamma)\), yielding maximum rates of around \(10^{2-3}\) events for an integrated luminosity of \(100\textrm{ fb}^{-1}\) in the mSUGRA scenario C' \cite{battaglia:2001zp}. The correlation is approximately independent of the neutrino parameters, but highly dependent on the mSUGRA parameters. As can be seen in Fig.~\ref{fig:br12_N2}~(right), sizeable rates of the decay \(\tilde\chi_2^0\to\tau^+\mu^-\tilde\chi_1^0\) are possible. In principle, this would make it feasible to test $\tau-\mu$ and $\tau-e$ flavour violation in the presence of the constraint on $Br(\mu\to e\gamma)$ in flavour symmetries, although it is necessary to update such analyses in light of new neutrino data, the updated limit on $Br(\mu\to e\gamma)$ and most importantly the recent LHC results. Analyses of LFV in generalized SUSY frameworks can be found in \cite{Bartl:2005yy, Carquin:2008gv}. In~\cite{Carquin:2008gv} it was concluded that the observation of the decay \(\tilde\chi_2^0 \to \tau^+\mu^-\tilde\chi_1^0\) will likely require a departure from strict mSUGRA boundary conditions and new sources of LFV such as given in SU(5) SUSY GUT models. 

\subsection{Low Scale Seesaw}
\label{sec:models_lss}

Without supersymmetry, lepton flavour violation can be enhanced by lowering the mass scale $M_R$ of the right-handed neutrinos while keeping the neutrino Yukawa couplings $Y^\nu$ reasonably large. In order to accommodate the observed mass scale of the light neutrinos $m^\nu = Y^{\nu T} M_R Y^\nu$, this either requires a fine-tuning between the structure of the Yukawa couplings and the Majorana mass terms, or a symmetry that conserves lepton number at $M_R$ and which is only broken by a small perturbation~\cite{Wyler:1982dd, Bernabeu:1987gr, Ilakovac:1994kj, Tommasini:1995ii, Pilaftsis:2004xx, Pilaftsis:2005rv} associated with a low scale of lepton number violation $m_\text{B-L}$. On general grounds, this leads to either decoupled or quasi-Dirac heavy neutrinos with mass splittings of the order $m_\text{B-L} / M_R$~\cite{Kersten:2007vk}. 
\begin{figure}[t]
\centering
\includegraphics[clip,width=0.65\textwidth]{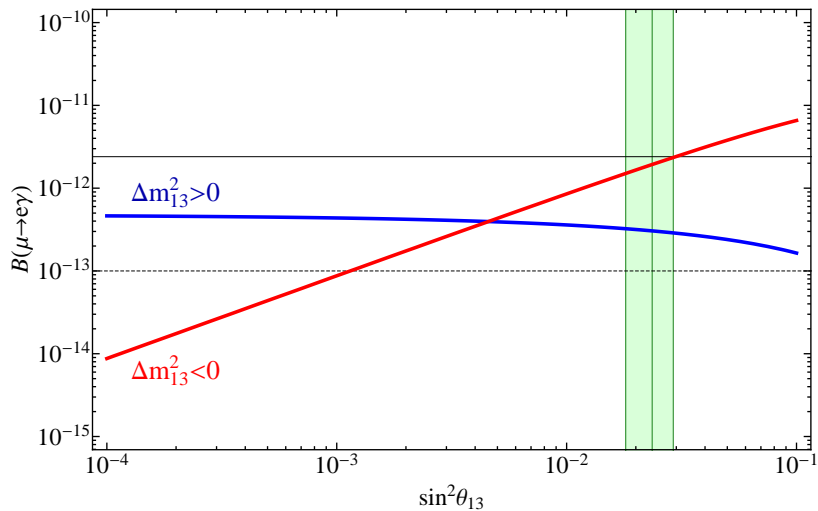}
\caption{Branching ratio $Br(\mu\to e\gamma)$ as a function of $\sin^2\theta_{13}$, in resonant leptogenesis scenarios with a normal (blue) and inverted (red) light neutrino mass spectrum (modified from \cite{Deppisch:2010fr}). The horizontal solid (dashed) line denotes the current (expected future) limit on $Br(\mu \to e\gamma)$. The vertical bar denotes the experimental $1\sigma$ range of $\sin^2\theta_{13}$~\cite{An:2012eh}.}  
\label{fig:theta13} 
\end{figure}
In ~\cite{Deppisch:2010fr}, such a scenario is realised by adopting the  lepton flavour symmetry ${\rm U(1)}_{L_e+L_\mu}\times {\rm U(1)}_{L_\tau}$, which is broken by small parameters $\epsilon_{e, \mu, \tau}$ and $\kappa_{1, 2}$,
\begin{equation}
	\label{eq:RLtau}
	Y^\nu =
	\begin{pmatrix}
		\epsilon_e    & a e^{-i\pi/4}  & a e^{i\pi/4} \\
		\epsilon_\mu  & b e^{-i\pi/4}  & b e^{i\pi/4} \\
		\epsilon_\tau & \kappa_1       & \kappa_2 
	\end{pmatrix}.
\end{equation}
In this model, the ${\rm U(1)}_{L_e+L_\mu}\times {\rm U(1)}_{L_\tau}$ invariant Yukawa couplings can be as large as $10^{-2}$, whereas the breaking parameters $\epsilon_i, \kappa_i$ are responsible for generating masses for two of the light neutrinos. This model was motivated to incorporate the resonant leptogenesis mechanism as an explanation of the observed baryon asymmetry of the universe~\cite{Pilaftsis:2004xx, Pilaftsis:2005rv}, where the lepton asymmetry in the decays of the heavy neutrinos is resonantly enhanced if the mass difference of the heavy neutrinos is of the order of their decay widths. This can be accommodated with the ansatz (\ref{eq:RLtau}) by radiatively inducing the small splitting of the heavy neutrinos with universal boundary conditions at the GUT scale~\cite{Deppisch:2010fr}, where the mass scale of the heavy neutrinos can be as low as 100~GeV.

The dominant Yukawa parameters $a,b \lesssim 10^{-2}$ induce potentially large $\mu-e$ flavour violation with
\begin{equation}
	Br(\mu\to e\gamma)\ \approx
	8.0\cdot 10^{-4}\times  \left(\frac{M_\text{EW}}{M_R}\right)^4 a^2 b^2,
\end{equation}
whereas $Br(\tau\to e\gamma)$ and $Br(\tau\to\mu\gamma)$ are small as they are suppressed by factors of $\kappa_i$. Taking into account all constraints from light neutrino data and the baryon asymmetry, the model is highly predictive and $a, b$ are not independent but can be expressed in terms of the light neutrino parameters. Fig.~\ref{fig:theta13} shows the resultant correlation of $Br(\mu\to e\gamma)$ with the light neutrino angle $\theta_{13}$ in two example scenarios with normal and inverted light neutrino mass hierarchies. 
The LFV rate is testable for large values of $\sin^2\theta_{13}$. The applied lepton flavour symmetry is a crucial ingredient for this enhancement as it allows large Yukawa couplings in addition to approximately protecting the $\tau$ lepton number for successful leptogenesis.

A similar enhancement of lepton flavour violating processes in the presence of a low scale of lepton number violation to lower the mass of heavy neutrinos has been demonstrated in \cite{Deppisch:2004fa, Deppisch:2005zm} for the so called inverse seesaw mechanism \cite{Bernabeu:1987gr}.

\subsection{Left-Right Symmetry}
\label{sec:models_lr}

In our final example, we will focus on the measurability of LFV couplings at the LHC in the minimal Left-Right symmetric model (LRSM). The following results are based on the analysis~\cite{Das:2012ii}. The Left-Right symmetrical model which extends the Standard Model gauge symmetry \321 to the \lr group~\cite{Pati:1974yy, Mohapatra:1974gc, Senjanovic:1975rk, Duka:1999uc}. Here, right-handed neutrinos are necessary to realise the extended gauge symmetry and come as part of an SU(2)$_R$ doublet, coupling to
the heavy gauge bosons. As a result, heavy neutrinos can be produced with gauge coupling strength, with promising discovery prospects.

\begin{figure}[t!]
\centering
\includegraphics[clip,width=0.32\textwidth]{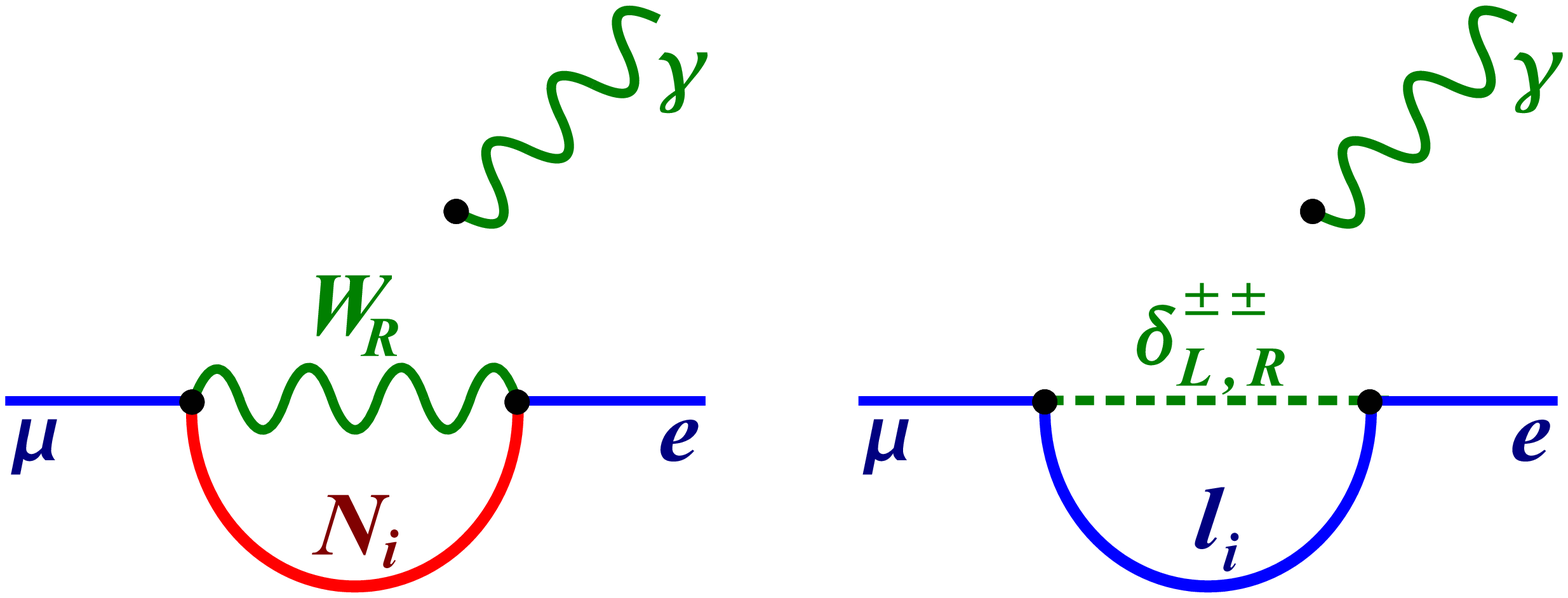}
\includegraphics[clip,width=0.32\textwidth]{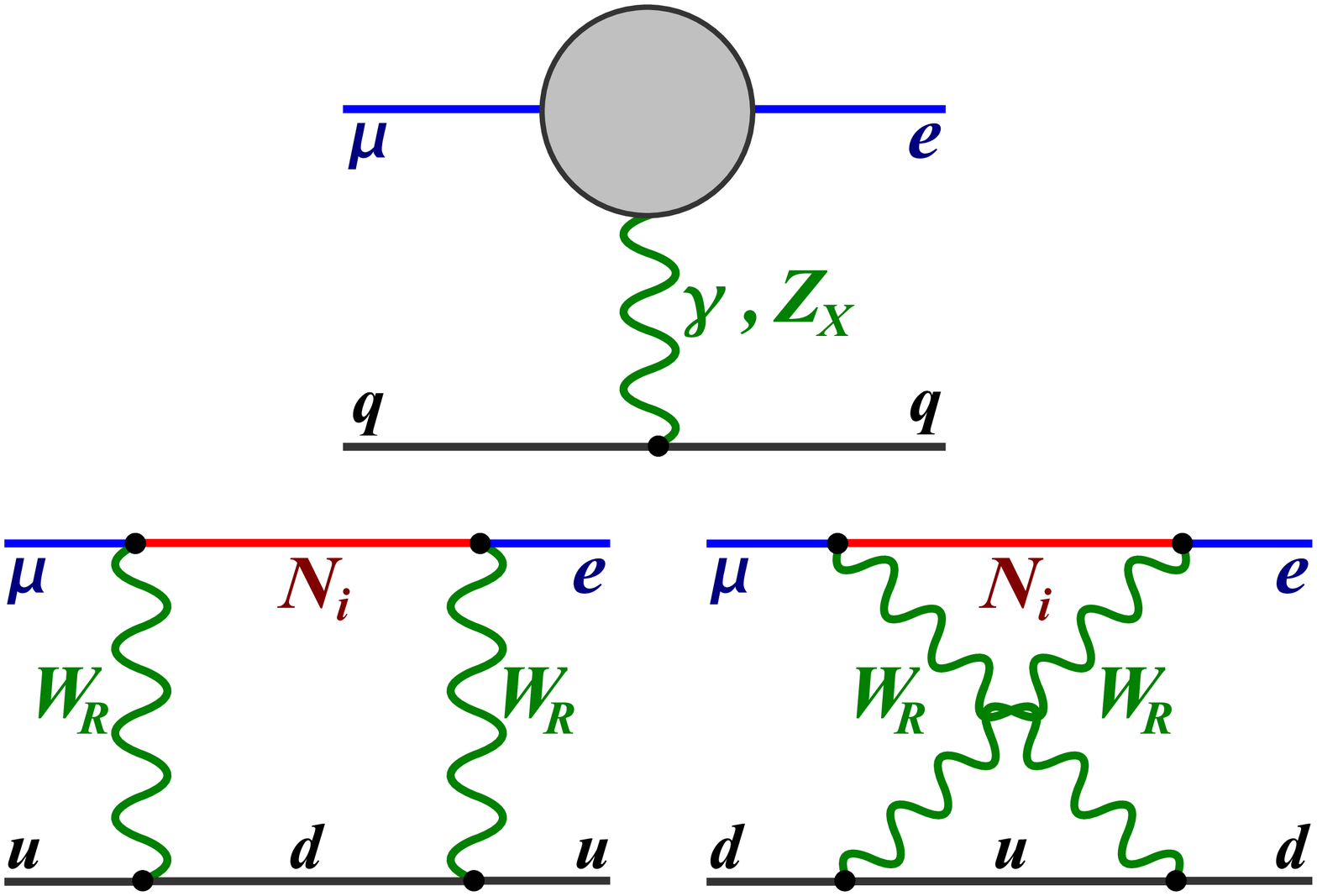}
\includegraphics[clip,width=0.32\textwidth]{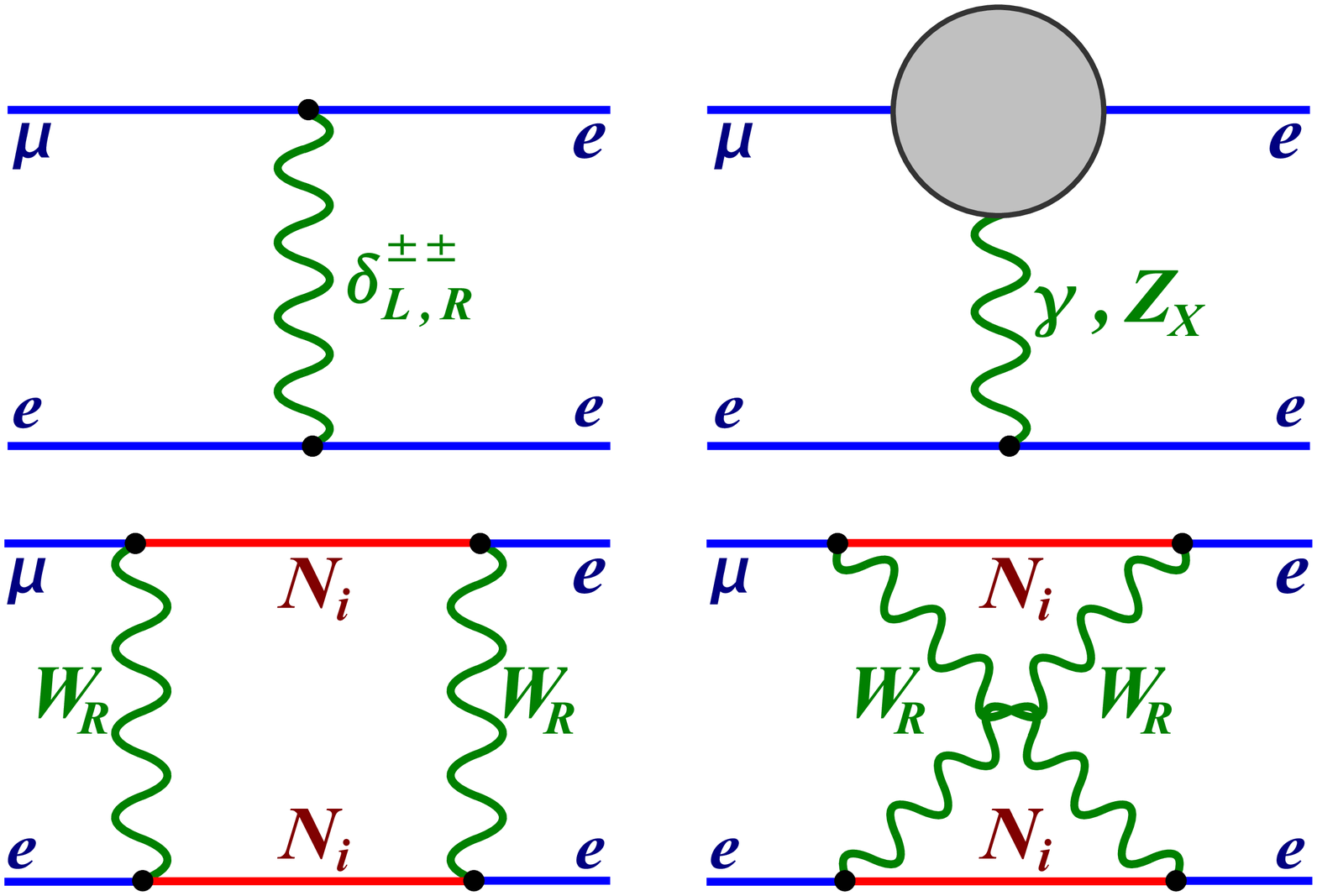}
\caption{Diagrams contributing to $\mu\to e\gamma$ (left), $\mu\to e$ conversion in nuclei (center) and $\mu\to eee$ (right) in left-right symmetry (from \cite{Das:2012ii}). The grey circle represents the effective $\mu-e-$gauge boson vertex of $\mu\to e\gamma$.}
\label{fig:diagrams_LFV_mue} 
\end{figure}
In the LRSM, a generation of leptons is assigned to the multiplet $L_i = (\nu_i, l_i)$ with the quantum numbers $Q_{L_L} = (1/2, 0, -1)$ and $Q_{L_R} = (0, 1/2, -1)$ under \lr. The Higgs sector contains a bidoublet $\phi$ and two triplets $\Delta_L$ and $\Delta_R$. The VEV $v_R$ of $\Delta_R$ breaks SU(2)$_R~\otimes$ U(1)$_{B-L}$ to U(1)$_Y$ and gives masses to a right-handed $W_R$ boson, a right-handed $Z_R$ boson and the heavy right-handed neutrinos. Since right-handed currents and particles have not been observed, $v_R$ should be sufficiently large. The VEVs of the neutral component of the bidoublet break the SM symmetry and are therefore of the order of the electroweak scale.

The LRSM can accommodate a general $6\times 6$ neutrino mass matrix in the basis $(\nu_L, \nu^c_L)^T$ of the form
\begin{equation}
	M =
	\begin{pmatrix}
		M_L & M_D \\ 
		M_D^T & M_R
	\end{pmatrix},
\label{eq:matr}
\end{equation}
with Majorana and Dirac mass entries of the order $M_L \approx y_M v_L$, $M_R \approx y_M v_R$ and $M_D=y_D m_\text{EW}$, where $y_{M,D}$ are Yukawa couplings. The Dirac mass term leads to a mixing between left- and right-handed neutrinos which is constrained by EW precision data to be $M_D/M_R \lesssim 10^{-2}$. The following results are reported in the regime of a dominant Seesaw I mechanism with a small Dirac mass term to accommodate the light neutrino masses $m_\nu = M_D^2/M_R$ with right-handed neutrino masses at the TeV scale. With $M_D \lesssim 10^{-4}$~GeV, an admixture of light and heavy neutrinos can be neglected. 

Fig.~\ref{fig:diagrams_LFV_mue} shows the diagrams contributing to the LFV processes $\mu\to e\gamma$, $\mu\to 3e$ and $\mu\to e$ conversion in nuclei, mediated by heavy neutrinos and doubly charged triplet Higgs bosons $\delta_{L,R}^{--}$. In general, branching ratios and conversion rates of these processes depend on many parameters, but under the assumption of similar mass scales between the heavy particles in the LRSM, $M_{N_i} \approx M_{W_R} \approx M_{\delta_{L,R}^{--}}$ one can make simple approximations. Such a spectrum is naturally expected, as all masses are generated in the breaking of the right-handed symmetry. In this approximation, the branching ratio of $\mu\to e\gamma$ is for example given by~\cite{Cirigliano:2004mv}
\begin{align}
\label{eq:BrmuegammaSimplified}
	Br(\mu\to e\gamma) 
	&\approx 1.5 \times 10^{-7} |g_{e\mu}|^2 \left(\frac{1\text{ TeV}}{M_{W_R}}\right)^4,
	\text{ with}\quad
	g_{e\mu} = 
		\sum_{n=1}^3 V^\dagger_{en} V^{\phantom{\dagger}}_{n\mu}
		\left(\frac{M_{N_n}}{M_{W_R}}\right)^2,
\end{align}
driven by the $3\times 3$ flavour mixing matrix $V$ of the right-handed neutrinos with the masses $M_{N_i}$, $i=1,2,3$. The lepton flavour violating processes have the following properties in the chosen scenario: (i) Both Br$(\mu\to e\gamma)$ and the $\mu-e$ conversion rate in nuclei $R_{\mu e}$ are proportional to the LFV factor $|g_{e\mu}|^2$. In addition, the ratio is $R_{\mu e}/Br(\mu\to e\gamma) = \mathcal{O}(1)$, independent of the right-handed neutrino mixing matrix $V$ and largely independent of the heavy particle spectrum. This is in stark contrast to models where the symmetry breaking occurs far above the electroweak scale, such as in supersymmetric seesaw models, where the photon penguin contribution dominates. (ii) Unless there are cancellations, one has $Br(\mu\to eee) / R_{\mu e} = \mathcal{O}(300)$ (for $M_{\delta_{L,R}^{--}} \approx$~1~TeV).

\begin{figure}[t]
\centering
\includegraphics[clip,width=0.49\textwidth]{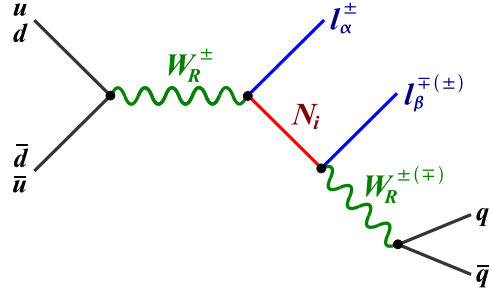}
\caption{Production and decay of a heavy right-handed neutrino with dilepton signature at hadron colliders (from \cite{Das:2012ii}).}
\label{fig:diagramsLHC}
\end{figure}
In \cite{Das:2012ii}, the LHC discovery potential of flavour violating dilepton signals $pp\to W_R \to e^\pm \mu^{\pm,\mp} +2$~jets via a heavy right-handed neutrino was assessed (cf. Fig.~\ref{fig:diagramsLHC}), with opposite sign (lepton number conserving) and same sign (lepton number violating) leptons in the final state. In the case of hierarchical right-handed neutrinos it is sufficient to consider one right-handed neutrino in the intermediate state. That is, either one right-handed neutrino is light enough to be produced, $M_{N_1} < M_{W_R} < M_{N_{2,3}}$, or for $M_{N_i} < M_{N_j} < M_{W_R}$, the mass difference between the right-handed neutrinos is sufficiently large such that the neutrino resonances can be individually reconstructed.

\begin{figure}[t]
\centering
\includegraphics[clip,width=0.49\textwidth]{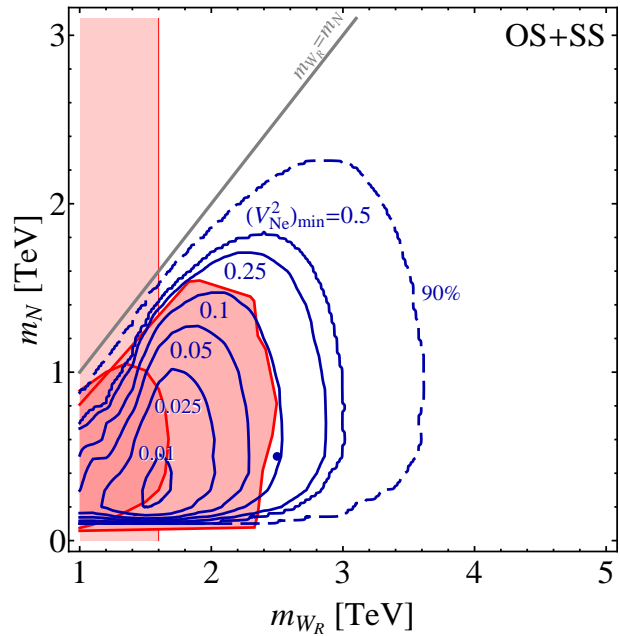}
\caption{Sensitivity to the coupling $|V_{Ne}|$ as function of $m_{W_R}$ and $m_{N_R}$ at the LHC with 14~TeV and $\mathcal{L}=30\text{ fb}^{-1}$ using both opposite and same sign lepton events (from \cite{Das:2012ii}). The solid contours indicate a discovery sensitivity at $5\sigma$ and the outermost dashed contour corresponds to an exclusion at 90\% C.L. for maximal mixing $|V_{Ne}|^2 = |V_{N\mu}|^2 = 1/2$. The shaded red areas are excluded by indirect (vertical bar) and direct LHC searches.}
\label{fig:Ve_MWMN} 
\end{figure}

Figure~\ref{fig:Ve_MWMN} shows the minimal coupling $|V_{Ne}|$ coupling of the heavy right-handed neutrino with an electron that can be observed at $5\sigma$. Here, a unitary mixing in the $e-\mu$ sector is assumed, $|V_{Ne}|^2 + |V_{N\mu}|^2 = 1$. Taking into account the direct exclusion limits from $W_R$ and $N_R$ LHC searches, flavour violating right-handed neutrino-lepton couplings down to $|V_{Ne(\mu)}| \approx 10^{-1}$ can potentially be probed at the LHC with 14~TeV and $\mathcal{L} = 30\text{ fb}^{-1}$. For LFV signals including a $\tau$ lepton an efficiency reduction by 30\% is expected~\cite{AguilarSaavedra:2012fu}. 

If two heavy neutrinos are light enough to be produced at the LHC, they will contribute to the dilepton LFV signature. For small mass differences this leads to interference effects and in the limit of degenerate heavy neutrinos, $\Delta M^2_{ij} \equiv M^2_{N_i}-M^2_{N_j} \to 0$, all lepton flavour violating signals will suffer a GIM-like suppression for unitary mixing. As a crucial difference to the radiative rare decays, the neutrinos with a short decay length are produced on-shell at the LHC. This leads to a decoherence of the right-handed neutrino oscillation, and the suppression is proportional to $\Delta M^2_{ij}/(M_N \Gamma_N)$, rather than $\Delta M^2_{ij}/M_N^2$. This follows from the well justified narrow width approximation of neutrino propagator~\cite{Deppisch:2003wt}.
\begin{figure}[t!]
\centering
\includegraphics[clip,width=0.467\textwidth]{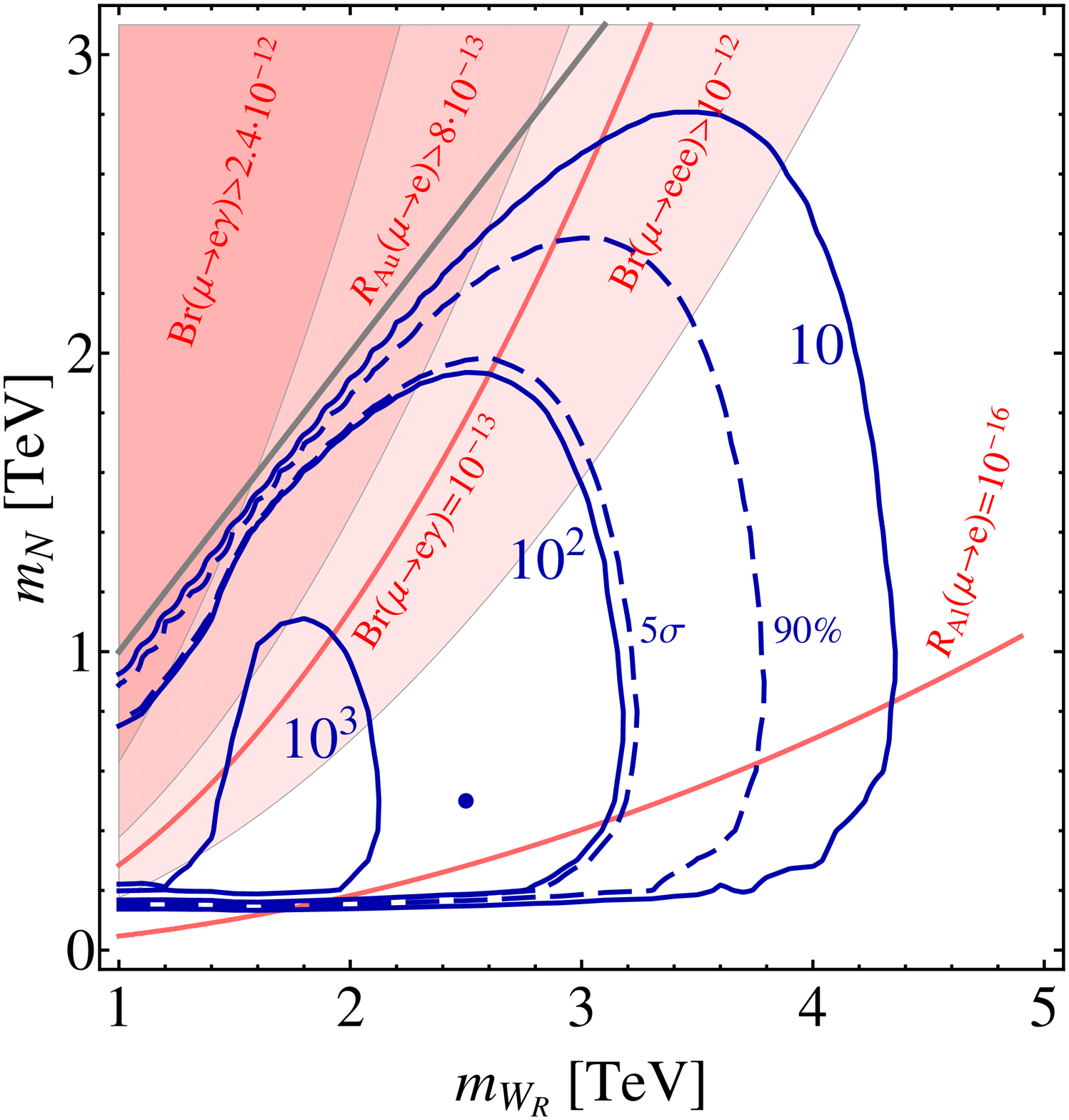}
\includegraphics[clip,width=0.513\textwidth]{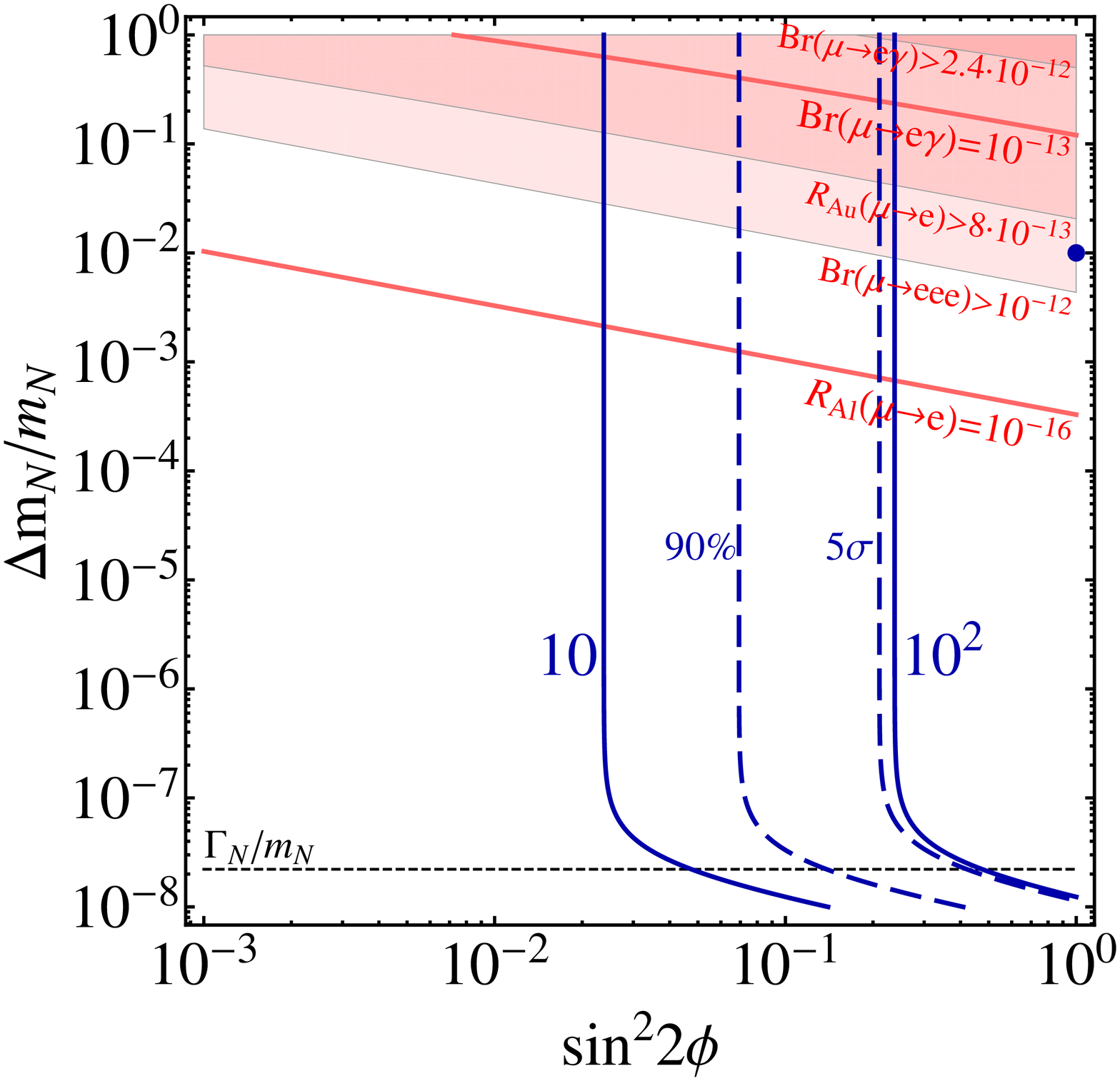}
\caption{
Dependence of the rates of low energy LFV processes (red contours and shaded areas) and the LFV signature $e^\pm\mu^{\pm,\mp} + 2j$ at the LHC (blue solid contours) with 14~TeV and $\mathcal{L}=30\text{ fb}^{-1}$ (from \cite{Das:2012ii}). The dashed contours define the parameter region with LHC signals at $5\sigma$ and 90\%, respectively. The processes have been calculated using maximal unitary mixing between two heavy neutrinos coupling only to $e$ and $\mu$. The spectrum of the doubly charged Higgs bosons is assumed to be $M_{\delta_{L,R}^{--}} = M_{W_R}$.
(Left) Dependence on the right-handed $W$ boson mass $M_{W_R}$ and the right-handed neutrino mass scale $M_N$ for maximal mixing and a 1\% mass splitting, $\phi=\pi/4$, $(M_{N_2} - M_{N_1})/M_N = 0.01$.
(Right) Dependence on the mixing angle parameter $\sin^2 2\phi$ and the heavy neutrino mass splitting $(M_{N_2}-M_{N_1})/M_N$ for $M_{W_R} = 2.5$~TeV and $M_N = 0.5$~TeV.}
\label{fig:ComparisonLowEnergy} 
\end{figure}
Fig.~\ref{fig:ComparisonLowEnergy} demonstrates this complementarity between collider and low-energy observables. It compares the sensitivity of $\mu-e$ LFV processes on the masses $M_{W_R}$ and $M_N$ for maximal unitary mixing $\phi = \pi/4$ between two right-handed neutrinos and a fixed 1\% neutrino mass splitting $(M_{N_2} - M_{N_1})/M_N = 10^{-2}$. This is shown in Fig.~\ref{fig:ComparisonLowEnergy}~(left). The current limits on the rare processes considerably constrain the parameter space, with $\mu\to eee$ providing the most stringent bound, due to the tree-level contribution from the doubly charged Higgs bosons shown in Fig.~\ref{fig:diagrams_LFV_mue}. On the other hand, Fig.~\ref{fig:ComparisonLowEnergy}~(right) shows the dependence on the right-handed neutrino mixing angle and mass difference for a fixed LRSM mass spectrum. As discussed above, the LFV process rate at the LHC is independent of the neutrino mass splitting until it becomes comparable to or smaller than the heavy neutrino decay width at $\Gamma_N / M_N \approx 5\cdot 10^{-8}$. It is therefore possible to probe such tiny mass splittings at the LHC for mixing angles $\sin^2(2\phi) \gsim 6\cdot 10^{-2}$ in this scenario. On the other hand, the low energy processes exhibit the typical GIM-suppressed dependence $\propto \sin^2(2\phi)(\Delta M_N^2)^2$, and may only probe mass splittings as low as $\Delta M_N / M_N \approx 10^{-3} - 10^{-4}$.

The above results demonstrate that under favourable conditions, LFV processes can be probed directly at the LHC, complementary to low-energy searches. Because the mediating particles are produced on-shell, the LHC has the potential to pin-point individual couplings. In the most optimistic scenario, all three neutrinos could be produced and if the mass differences are large enough, could be individually reconstructed and their coupling strengths measured. Although challenging, this could also include LFV signatures containing $\tau$ leptons with an efficiency hit of 30\%~\cite{AguilarSaavedra:2012fu}. Specifically, at the time of writing, it is still possible to measure large right-handed neutrino mixing strengths $V_{Nl} = 0.1 - 1$, but searches at the LHC are progressing fast and by the end of 2012 it should be clear if this is still a viable scenario.

\section{Summary}
\label{sec:summary}

The observation of neutrino oscillations has confirmed the presence of flavour violating effects in leptons. Because of the extremely small neutrino masses and mass differences, generally understood to be originating from the breaking of lepton number at a high scale, the induced flavour violation in charged leptons is naturally suppressed to be unobservably small. This could be seen as another success of the Standard Model (incorporating light neutrino masses), as no charged lepton flavour violating processes have been observed so far, despite impressive experimental efforts pushing the sensitivity by currently excluding rates as low as $1$~LFV event among $10^{12}$~standard events.

On the other hand, many frameworks beyond the Standard Model predict new flavour violating sources inducing potentially large and experimentally excluded charged LFV process rates. This often leads to the formulation of a flavour problem in BSM frameworks, such as in supersymmetry, identifying the need for a mechanism to suppress large flavour violating operators. The incorporation of a weakly broken flavour symmetry among the fermion generations in order to suppress flavour changing interactions can in general provide such a mechanism.
In this article we have reviewed the implications of models incorporating flavour symmetries on charged lepton flavour violating processes. The non-observation of processes such as $\mu\to e\gamma$ already puts stringent constraints on models of new physics. If observed, they can provide crucial information to pinpoint the lepton mixing properties and help solve the flavour puzzle.

As examples of new physics frameworks we have focused on scenarios that incorporate a  Seesaw mechanism to generate the light neutrino masses, either with or without supersymmetry and with right-handed neutrinos close to the GUT scale or the electroweak scale. In this regard, we would like to stress that there is a host of other BSM frameworks in which lepton flavour symmetries can be embedded, such as R-parity violating SUSY, Extra-dimensional models, Little Higgs models or models with additional light SU(2) Higgs doublets.

With the high sensitivity of searches for low energy LFV processes, such as $\mu\to e\gamma$ and $\mu\to e$~conversion in nuclei, they currently provide the most stringent constraints on new physics of lepton flavour violation. Flavour symmetries generally predict relative LFV rates in different flavour transitions, e.g. $Br(\mu\to e\gamma) : Br(\tau\to \mu\gamma) : Br(\tau\to e\gamma)$, whereas the total rate is suppressed by the scale where new physics mediating the LFV effects enters. Because of the large difference in the experimental sensitivity between LFV decays of muons and taus, $Br_\text{exp. limit}(\mu\to e\gamma) : Br_\text{exp. limit}(\tau\to \ell\gamma) \approx 10^{-4}$, it is sometimes difficult to connect predictions of flavour symmetries with phenomenology. Observations at the LHC, and high energy colliders in general, can provide crucial information in two ways. Firstly, and most obviously, if new physics is observed at the LHC, the scale at which it enters can be fixed and correlated with the total rate of LFV processes. Secondly, if the actual LFV mediating particles can be produced, their couplings can potentially be probed in high detail, providing complementary information to low energy LFV observables that may help to pinpoint the fermion mixing structure and solve the flavour puzzle of physics.


\bibliographystyle{h-physrev4}
\bibliography{FrankDeppisch}

\begin{thebibliography}{100}

\bibitem{fukuda:1998mi}
Super-Kamiokande collaboration, Y.~Fukuda {\em et~al.},
\newblock Phys. Rev. Lett. {\bf 81}, 1562 (1998), [hep-ex/9807003].

\bibitem{ahmad:2002jz}
SNO collaboration, Q.~R. Ahmad {\em et~al.},
\newblock Phys. Rev. Lett. {\bf 89}, 011301 (2002), [nucl-ex/0204008].

\bibitem{eguchi:2002dm}
KamLAND collaboration, K.~Eguchi {\em et~al.},
\newblock Phys. Rev. Lett. {\bf 90}, 021802 (2003), [hep-ex/0212021].

\bibitem{nunokawa:2007qh}
H.~Nunokawa, S.~J. Parke and J.~W.~F. Valle,
\newblock Prog. Part. Nucl. Phys. {\bf 60}, 338 (2008).

\bibitem{Abada:2011rg}
A.~Abada,
\newblock Comptes Rendus Physique {\bf 13}, 180 (2012), [1110.6507],
\newblock 8 pages, 6 figures/ to appear in C. R. Physique (2011).

\bibitem{Maltoni:2004ei}
M.~Maltoni, T.~Schwetz, M.~A. Tortola and J.~W.~F. Valle,
\newblock New J. Phys. {\bf 6}, 122 (2004),
\newblock hep-ph/0405172.

\bibitem{Bhattacharyya:2010hp}
G.~Bhattacharyya, P.~Leser and H.~P{\"a}s,
\newblock Phys.Rev. {\bf D83}, 011701 (2011), [1006.5597].

\bibitem{Joshipura:1990pi}
A.~S. Joshipura and S.~D. Rindani,
\newblock Phys.Lett. {\bf B260}, 149 (1991).

\bibitem{Branco:1996bq}
G.~Branco, W.~Grimus and L.~Lavoura,
\newblock Phys.Lett. {\bf B380}, 119 (1996), [hep-ph/9601383].

\bibitem{Kubo:2003iw}
J.~Kubo, A.~Mondragon, M.~Mondragon and E.~Rodriguez-Jauregui,
\newblock Prog. Theor. Phys. {\bf 109}, 795 (2003), [hep-ph/0302196].

\bibitem{Botella:2009pq}
F.~Botella, G.~Branco and M.~Rebelo,
\newblock Phys.Lett. {\bf B687}, 194 (2010), [0911.1753].

\bibitem{Froggatt:1978nt}
C.~Froggatt and H.~B. Nielsen,
\newblock Nucl.Phys. {\bf B147}, 277 (1979).

\bibitem{Barbieri:1996ae}
R.~Barbieri and L.~J. Hall,
\newblock Nuovo Cim. {\bf A110}, 1 (1997), [hep-ph/9605224].

\bibitem{Barbieri:1996ww}
R.~Barbieri, L.~J. Hall, S.~Raby and A.~Romanino,
\newblock Nucl.Phys. {\bf B493}, 3 (1997), [hep-ph/9610449].

\bibitem{Barbieri:1997tu}
R.~Barbieri, L.~J. Hall and A.~Romanino,
\newblock Phys.Lett. {\bf B401}, 47 (1997), [hep-ph/9702315].

\bibitem{King:2005bj}
S.~King,
\newblock JHEP {\bf 0508}, 105 (2005), [hep-ph/0506297].

\bibitem{King:2001uz}
S.~King and G.~G. Ross,
\newblock Phys.Lett. {\bf B520}, 243 (2001), [hep-ph/0108112].

\bibitem{King:2003rf}
S.~King and G.~G. Ross,
\newblock Phys.Lett. {\bf B574}, 239 (2003), [hep-ph/0307190],
\newblock Dedicated to Ian I. Kogan.

\bibitem{Harrison:2002er}
P.~Harrison, D.~Perkins and W.~Scott,
\newblock Phys.Lett. {\bf B530}, 167 (2002), [hep-ph/0202074].

\bibitem{He:2003rm}
X.~G. He and A.~Zee,
\newblock Phys.Lett. {\bf B560}, 87 (2003), [hep-ph/0301092].

\bibitem{Altarelli:2005yx}
G.~Altarelli and F.~Feruglio,
\newblock Nucl.Phys. {\bf B741}, 215 (2006), [hep-ph/0512103].

\bibitem{An:2012eh}
DAYA-BAY Collaboration, F.~An {\em et~al.},
\newblock Phys.Rev.Lett. {\bf 108}, 171803 (2012), [1203.1669],
\newblock 5 figures. Version to appear in Phys. Rev. Lett.

\bibitem{Ahn:2012nd}
RENO collaboration, J.~Ahn {\em et~al.},
\newblock 1204.0626.

\bibitem{Adhikary:2008au}
B.~Adhikary and A.~Ghosal,
\newblock Phys.Rev. {\bf D78}, 073007 (2008), [0803.3582].

\bibitem{Ishimori:2012fg}
H.~Ishimori and E.~Ma,
\newblock 1205.0075.

\bibitem{Leurer:1992wg}
M.~Leurer, Y.~Nir and N.~Seiberg,
\newblock Nucl.Phys. {\bf B398}, 319 (1993), [hep-ph/9212278].

\bibitem{Dine:1993np}
M.~Dine, R.~G. Leigh and A.~Kagan,
\newblock Phys.Rev. {\bf D48}, 4269 (1993), [hep-ph/9304299].

\bibitem{Nir:1993mx}
Y.~Nir and N.~Seiberg,
\newblock Phys.Lett. {\bf B309}, 337 (1993), [hep-ph/9304307].

\bibitem{Leurer:1993gy}
M.~Leurer, Y.~Nir and N.~Seiberg,
\newblock Nucl.Phys. {\bf B420}, 468 (1994), [hep-ph/9310320].

\bibitem{Pomarol:1995xc}
A.~Pomarol and D.~Tommasini,
\newblock Nucl.Phys. {\bf B466}, 3 (1996), [hep-ph/9507462].

\bibitem{Barbieri:1998qs}
R.~Barbieri, L.~J. Hall and A.~Romanino,
\newblock Nucl.Phys. {\bf B551}, 93 (1999), [hep-ph/9812384].

\bibitem{Berezhiani:2000cg}
Z.~Berezhiani and A.~Rossi,
\newblock Nucl.Phys. {\bf B594}, 113 (2001), [hep-ph/0003084].

\bibitem{Aranda:2000tm}
A.~Aranda, C.~D. Carone and R.~F. Lebed,
\newblock Phys. Rev. {\bf D62}, 016009 (2000), [hep-ph/0002044].

\bibitem{Lavignac:2001vp}
S.~Lavignac, I.~Masina and C.~A. Savoy,
\newblock Phys. Lett. {\bf B520}, 269 (2001), [hep-ph/0106245].

\bibitem{Roberts:2001zy}
R.~Roberts, A.~Romanino, G.~G. Ross and L.~Velasco-Sevilla,
\newblock Nucl.Phys. {\bf B615}, 358 (2001), [hep-ph/0104088].

\bibitem{Ross:2002fb}
G.~G. Ross and L.~Velasco-Sevilla,
\newblock Nucl. Phys. {\bf B653}, 3 (2003), [hep-ph/0208218].

\bibitem{Nir:2002ah}
Y.~Nir and G.~Raz,
\newblock Phys.Rev. {\bf D66}, 035007 (2002), [hep-ph/0206064].

\bibitem{Kane:2005va}
G.~Kane, S.~King, I.~Peddie and L.~Velasco-Sevilla,
\newblock JHEP {\bf 0508}, 083 (2005), [hep-ph/0504038].

\bibitem{Hirsch:2012ym}
M.~Hirsch {\em et~al.},
\newblock 1201.5525,
\newblock Long author list - awaiting processing.

\bibitem{Carone:1995xw}
C.~D. Carone, L.~J. Hall and H.~Murayama,
\newblock Phys.Rev. {\bf D53}, 6282 (1996), [hep-ph/9512399].

\bibitem{Nir:1996am}
Y.~Nir and R.~Rattazzi,
\newblock Phys.Lett. {\bf B382}, 363 (1996), [hep-ph/9603233].

\bibitem{Binetruy:1996xk}
P.~Binetruy, S.~Lavignac and P.~Ramond,
\newblock Nucl. Phys. {\bf B477}, 353 (1996), [hep-ph/9601243].

\bibitem{Dudas:1996fe}
E.~Dudas, C.~Grojean, S.~Pokorski and C.~A. Savoy,
\newblock Nucl.Phys. {\bf B481}, 85 (1996), [hep-ph/9606383].

\bibitem{Carone:1997qg}
C.~D. Carone and L.~J. Hall,
\newblock Phys.Rev. {\bf D56}, 4198 (1997), [hep-ph/9702430],
\newblock 23 pp. LaTeX Report-no: LBNL-40024, UCB-PTH-97/08.

\bibitem{Barbieri:1998em}
R.~Barbieri, L.~Giusti, L.~J. Hall and A.~Romanino,
\newblock Nucl.Phys. {\bf B550}, 32 (1999), [hep-ph/9812239].

\bibitem{Hall:1998cu}
L.~J. Hall and N.~Weiner,
\newblock Phys.Rev. {\bf D60}, 033005 (1999), [hep-ph/9811299].

\bibitem{Aranda:1999kc}
A.~Aranda, C.~D. Carone and R.~F. Lebed,
\newblock Phys.Lett. {\bf B474}, 170 (2000), [hep-ph/9910392].

\bibitem{Barbieri:1999pe}
R.~Barbieri, P.~Creminelli and A.~Romanino,
\newblock Nucl.Phys. {\bf B559}, 17 (1999), [hep-ph/9903460].

\bibitem{Aranda:2001rd}
A.~Aranda, C.~D. Carone and P.~Meade,
\newblock Phys.Rev. {\bf D65}, 013011 (2002), [hep-ph/0109120].

\bibitem{Chen:2000fp}
M.-C. Chen and K.~Mahanthappa,
\newblock Phys.Rev. {\bf D62}, 113007 (2000), [hep-ph/0005292].

\bibitem{Chkareuli:2001dq}
J.~Chkareuli, C.~Froggatt and H.~Nielsen,
\newblock Nucl.Phys. {\bf B626}, 307 (2002), [hep-ph/0109156].

\bibitem{Dreiner:2003yr}
H.~K. Dreiner, H.~Murayama and M.~Thormeier,
\newblock Nucl.Phys. {\bf B729}, 278 (2005), [hep-ph/0312012].

\bibitem{Ross:2004qn}
G.~G. Ross, L.~Velasco-Sevilla and O.~Vives,
\newblock Nucl.Phys. {\bf B692}, 50 (2004), [hep-ph/0401064].

\bibitem{Kaplan:1993ej}
D.~B. Kaplan and M.~Schmaltz,
\newblock Phys.Rev. {\bf D49}, 3741 (1994), [hep-ph/9311281].

\bibitem{Ma:2001dn}
E.~Ma and G.~Rajasekaran,
\newblock Phys. Rev. {\bf D64}, 113012 (2001), [hep-ph/0106291].

\bibitem{Babu:2002dz}
K.~S. Babu, E.~Ma and J.~W.~F. Valle,
\newblock Phys. Lett. {\bf B552}, 207 (2003), [hep-ph/0206292].

\bibitem{Grimus:2004rj}
W.~Grimus, A.~S. Joshipura, S.~Kaneko, L.~Lavoura and M.~Tanimoto,
\newblock JHEP {\bf 07}, 078 (2004), [hep-ph/0407112].

\bibitem{Grimus:2005mu}
W.~Grimus and L.~Lavoura,
\newblock JHEP {\bf 0508}, 013 (2005), [hep-ph/0504153].

\bibitem{Hagedorn:2006ug}
C.~Hagedorn, M.~Lindner and R.~Mohapatra,
\newblock JHEP {\bf 0606}, 042 (2006), [hep-ph/0602244].

\bibitem{King:2006np}
S.~F. King and M.~Malinsky,
\newblock Phys. Lett. {\bf B645}, 351 (2007), [hep-ph/0610250].

\bibitem{Feruglio:2007uu}
F.~Feruglio, C.~Hagedorn, Y.~Lin and L.~Merlo,
\newblock Nucl. Phys. {\bf B775}, 120 (2007), [hep-ph/0702194].

\bibitem{Kubo:2004ps}
J.~Kubo, H.~Okada and F.~Sakamaki,
\newblock Phys.Rev. {\bf D70}, 036007 (2004), [hep-ph/0402089].

\bibitem{Chen:2004rr}
S.-L. Chen, M.~Frigerio and E.~Ma,
\newblock Phys.Rev. {\bf D70}, 073008 (2004), [hep-ph/0404084].

\bibitem{Lavoura:2005kx}
L.~Lavoura and E.~Ma,
\newblock Mod.Phys.Lett. {\bf A20}, 1217 (2005), [hep-ph/0502181].

\bibitem{Teshima:2005bk}
T.~Teshima,
\newblock Phys.Rev. {\bf D73}, 045019 (2006), [hep-ph/0509094].

\bibitem{Koide:2005ep}
Y.~Koide,
\newblock Phys.Rev. {\bf D73}, 057901 (2006), [hep-ph/0509214].

\bibitem{Mohapatra:2006pu}
R.~Mohapatra, S.~Nasri and H.-B. Yu,
\newblock Phys.Lett. {\bf B639}, 318 (2006), [hep-ph/0605020].

\bibitem{Morisi:2005fy}
S.~Morisi and M.~Picariello,
\newblock Int.J.Theor.Phys. {\bf 45}, 1267 (2006), [hep-ph/0505113].

\bibitem{Kaneko:2006wi}
S.~Kaneko, H.~Sawanaka, T.~Shingai, M.~Tanimoto and K.~Yoshioka,
\newblock Prog.Theor.Phys. {\bf 117}, 161 (2007), [hep-ph/0609220].

\bibitem{Bazzocchi:2007au}
F.~Bazzocchi, S.~Morisi and M.~Picariello,
\newblock Phys.Lett. {\bf B659}, 628 (2008), [0710.2928].

\bibitem{Hirsch:2008mg}
M.~Hirsch, S.~Morisi and J.~W.~F. Valle,
\newblock Phys. Rev. {\bf D79}, 016001 (2009), [0810.0121].

\bibitem{Hagedorn:2008bc}
C.~Hagedorn, M.~A. Schmidt and A.~Y. Smirnov,
\newblock Phys.Rev. {\bf D79}, 036002 (2009), [0811.2955].

\bibitem{Feruglio:2009iu}
F.~Feruglio, C.~Hagedorn and L.~Merlo,
\newblock JHEP {\bf 1003}, 084 (2010), [0910.4058].

\bibitem{Bazzocchi:2009pv}
F.~Bazzocchi, L.~Merlo and S.~Morisi,
\newblock Nucl.Phys. {\bf B816}, 204 (2009), [0901.2086].

\bibitem{Bazzocchi:2009da}
F.~Bazzocchi, L.~Merlo and S.~Morisi,
\newblock Phys. Rev. {\bf D80}, 053003 (2009), [0902.2849].

\bibitem{Hirsch:2009mx}
M.~Hirsch, S.~Morisi and J.~W.~F. Valle,
\newblock Phys. Lett. {\bf B679}, 454 (2009), [0905.3056].

\bibitem{Morisi:2009sc}
S.~Morisi and E.~Peinado,
\newblock Phys. Rev. {\bf D80}, 113011 (2009), [0910.4389].

\bibitem{Hirsch:2010ru}
M.~Hirsch, S.~Morisi, E.~Peinado and J.~Valle,
\newblock Phys.Rev. {\bf D82}, 116003 (2010), [1007.0871].

\bibitem{Hagedorn:2010th}
C.~Hagedorn, S.~F. King and C.~Luhn,
\newblock JHEP {\bf 1006}, 048 (2010), [1003.4249].

\bibitem{Hagedorn:2010mq}
C.~Hagedorn and R.~Ziegler,
\newblock Phys.Rev. {\bf D82}, 053011 (2010), [1007.1888].

\bibitem{Boucenna:2011tj}
M.~Boucenna {\em et~al.},
\newblock JHEP {\bf 1105}, 037 (2011), [1101.2874].

\bibitem{Toorop:2011jn}
R.~d.~A. Toorop, F.~Feruglio and C.~Hagedorn,
\newblock Phys.Lett. {\bf B703}, 447 (2011), [1107.3486],
\newblock 1+11 pages, 1 figure/ v2: matches journal version.

\bibitem{deAdelhartToorop:2011ad}
R.~de~Adelhart~Toorop, F.~Bazzocchi and S.~Morisi,
\newblock Nucl.Phys. {\bf B856}, 670 (2012), [1104.5676].

\bibitem{Hagedorn:2012pg}
C.~Hagedorn and D.~Meloni,
\newblock 1204.0715.

\bibitem{Hoecker:2012nu}
A.~Hoecker,
\newblock 1201.5093.

\bibitem{Feldmann:2011zh}
T.~Feldmann,
\newblock PoS {\bf BEAUTY2011}, 017 (2011), [1105.2139].

\bibitem{Adam:2009ci}
MEG collaboration, J.~Adam {\em et~al.},
\newblock Nucl.Phys. {\bf B834}, 1 (2010), [0908.2594].

\bibitem{Adam:2011ch}
MEG collaboration, J.~Adam {\em et~al.},
\newblock Phys.Rev.Lett. {\bf 107}, 171801 (2011), [1107.5547],
\newblock 5 pages, 2 figures, accepted for publication at Phys. Rev. Lett.

\bibitem{Bellgardt:1987du}
SINDRUM Collaboration, U.~Bellgardt {\em et~al.},
\newblock Nucl.Phys. {\bf B299}, 1 (1988).

\bibitem{Hewett:2012IntensityFrontier}
R. C. Group, J.~Hewett {\em et~al.},
\newblock 1205.2671,
\newblock 229 pages.

\bibitem{Deppisch:2005zm}
F.~Deppisch, T.~Kosmas and J.~Valle,
\newblock Nucl.Phys. {\bf B752}, 80 (2006), [hep-ph/0512360].

\bibitem{Czarnecki:1998iz}
A.~Czarnecki, W.~J. Marciano and K.~Melnikov,
\newblock AIP Conf.Proc. {\bf 435}, 409 (1998), [hep-ph/9801218].

\bibitem{Kitano:2002mt}
R.~Kitano, M.~Koike and Y.~Okada,
\newblock Phys. Rev. {\bf D66}, 096002 (2002), [hep-ph/0203110].

\bibitem{Marciano:2008zz}
W.~J. Marciano, T.~Mori and J.~M. Roney,
\newblock Ann.Rev.Nucl.Part.Sci. {\bf 58}, 315 (2008).

\bibitem{Raidal:2008jk}
M.~Raidal {\em et~al.},
\newblock Eur.Phys.J. {\bf C57}, 13 (2008), [0801.1826].

\bibitem{Cirigliano:2009bz}
V.~Cirigliano, R.~Kitano, Y.~Okada and P.~Tuzon,
\newblock Phys.Rev. {\bf D80}, 013002 (2009), [0904.0957].

\bibitem{Bertl:2006up}
SINDRUM II Collaboration, W.~H. Bertl {\em et~al.},
\newblock Eur.Phys.J. {\bf C47}, 337 (2006).

\bibitem{Kutschke:2011ux}
R.~K. Kutschke,
\newblock 1112.0242.

\bibitem{Kurup:2011zza}
COMET Collaboration, A.~Kurup,
\newblock Nucl.Phys.Proc.Suppl. {\bf 218}, 38 (2011).

\bibitem{Tschirhart:2011zza}
R.~Tschirhart,
\newblock Nucl.Phys.Proc.Suppl. {\bf 210-211}, 203 (2011).

\bibitem{Barlow:2011zz}
R.~Barlow,
\newblock Nucl.Phys.Proc.Suppl. {\bf 218}, 44 (2011).

\bibitem{Asner:2010qj}
Heavy Flavor Averaging Group, D.~Asner {\em et~al.},
\newblock 1010.1589,
\newblock Long author list - awaiting processing.

\bibitem{Aubert:2009ag}
BABAR Collaboration, B.~Aubert {\em et~al.},
\newblock Phys.Rev.Lett. {\bf 104}, 021802 (2010), [0908.2381],
\newblock 7 pages, 2 encapsulated postscript figures, submitted to Physical
  Review Letters.

\bibitem{Miyazaki:2011xe}
Belle Collaboration, Y.~Miyazaki {\em et~al.},
\newblock Phys.Lett. {\bf B699}, 251 (2011), [1101.0755],
\newblock Long author list - awaiting processing.

\bibitem{Bona:2007qt}
SuperB Collaboration, M.~Bona {\em et~al.},
\newblock 0709.0451.

\bibitem{Abe:2010sj}
T.~A. et~al. (Belle~II),
\newblock arXiv/1011.0352.

\bibitem{1742-6596-171-1-012079}
K.~Hayasaka,
\newblock Journal of Physics: Conference Series {\bf 171}, 012079 (2009).

\bibitem{Giffels:2008ar}
M.~Giffels, J.~Kallarackal, M.~Kramer, B.~O'Leary and A.~Stahl,
\newblock Phys.Rev. {\bf D77}, 073010 (2008), [0802.0049].

\bibitem{Seyfert:1451298}
P.~Seyfert,
\newblock (2012),
\newblock Talk given at Flavour Physics and CP Violation 2012, Hefei, Anhui,
  China, 21 - 25 May 2012.

\bibitem{LHCb-CONF-2012-015}
J.~Albrecht and et~al. (LHCb),
\newblock (2012),
\newblock Linked to LHCb-ANA-2011-100.

\bibitem{AguilarSaavedra:2012fu}
J.~Aguilar-Saavedra, F.~Deppisch, O.~Kittel and J.~Valle,
\newblock Phys.Rev. {\bf D85}, 091301 (2012), [1203.5998].

\bibitem{Altarelli:2010gt}
G.~Altarelli and F.~Feruglio,
\newblock Rev.Mod.Phys. {\bf 82}, 2701 (2010), [1002.0211].

\bibitem{Calibbi:2012at}
L.~Calibbi, Z.~Lalak, S.~Pokorski and R.~Ziegler,
\newblock 1204.1275.

\bibitem{Merlo:2010mw}
L.~Merlo,
\newblock 1004.2211.

\bibitem{Feruglio:2009hu}
F.~Feruglio, C.~Hagedorn, Y.~Lin and L.~Merlo,
\newblock Nucl.Phys. {\bf B832}, 251 (2010), [0911.3874].

\bibitem{Hagedorn:2009df}
C.~Hagedorn, E.~Molinaro and S.~Petcov,
\newblock JHEP {\bf 1002}, 047 (2010), [0911.3605].

\bibitem{gell-mann:1980vs}
M.~Gell-Mann, P.~Ramond and R.~Slansky,
\newblock (1979),
\newblock Print-80-0576 (CERN).

\bibitem{Minkowski:1977sc}
P.~Minkowski,
\newblock Phys. Lett. {\bf B67}, 421 (1977).

\bibitem{Glashow:1979}
S.~Glashow,
\newblock (1980),
\newblock ed. M. Levy et al. (Plenum, New York), p. 707.

\bibitem{Mohapatra:1979ia}
R.~N. Mohapatra and G.~Senjanovic,
\newblock Phys. Rev. Lett. {\bf 44}, 912 (1980).

\bibitem{Yanagida:1979}
T.~Yanagida,
\newblock (KEK lectures, 1979),
\newblock ed. O. Sawada and A. Sugamoto (KEK, 1979).

\bibitem{Schechter:1980gr}
J.~Schechter and J.~W.~F. Valle,
\newblock Phys. Rev. {\bf D22}, 2227 (1980).

\bibitem{Mohapatra:1981yp}
R.~N. Mohapatra and G.~Senjanovic,
\newblock Phys. Rev. {\bf D23}, 165 (1981).

\bibitem{Hall:1986dx}
L.~J. Hall, V.~A. Kostelecky and S.~Raby,
\newblock Nucl. Phys. {\bf B267}, 415 (1986).

\bibitem{Borzumati:1986qx}
F.~Borzumati and A.~Masiero,
\newblock Phys. Rev. Lett. {\bf 57}, 961 (1986).

\bibitem{Barbieri:1994pv}
R.~Barbieri and L.~J. Hall,
\newblock Phys. Lett. {\bf B338}, 212 (1994), [hep-ph/9408406].

\bibitem{Hisano:1996cp}
J.~Hisano, T.~Moroi, K.~Tobe and M.~Yamaguchi,
\newblock Phys. Rev. {\bf D53}, 2442 (1996), [hep-ph/9510309].

\bibitem{Hisano:1999fj}
J.~Hisano and D.~Nomura,
\newblock Phys. Rev. {\bf D59}, 116005 (1999), [hep-ph/9810479].

\bibitem{Casas:2001sr}
J.~A. Casas and A.~Ibarra,
\newblock Nucl. Phys. {\bf B618}, 171 (2001), [hep-ph/0103065].

\bibitem{Kageyama:2001tn}
A.~Kageyama, S.~Kaneko, N.~Shimoyama and M.~Tanimoto,
\newblock Phys.Rev. {\bf D65}, 096010 (2002), [hep-ph/0112359].

\bibitem{Deppisch:2002vz}
F.~Deppisch, H.~Paes, A.~Redelbach, R.~R{\"u}ckl and Y.~Shimizu,
\newblock Eur. Phys. J. {\bf C28}, 365 (2003), [hep-ph/0206122].

\bibitem{Deppisch:2002tv}
F.~Deppisch, H.~P{\"a}s, A.~Redelbach, R.~R{\"u}ckl and Y.~Shimizu,
\newblock hep-ph/0210407.

\bibitem{Deppisch:2002qw}
F.~Deppisch, H.~P{\"a}s, A.~Redelbach, R.~R{\"u}ckl and Y.~Shimizu,
\newblock Nucl.Phys.Proc.Suppl. {\bf 116}, 316 (2003), [hep-ph/0211138].

\bibitem{Chivukula:1987py}
R.~S. Chivukula and H.~Georgi,
\newblock Phys.Lett. {\bf B188}, 99 (1987).

\bibitem{Hall:1990ac}
L.~Hall and L.~Randall,
\newblock Phys.Rev.Lett. {\bf 65}, 2939 (1990).

\bibitem{D'Ambrosio:2002ex}
G.~D'Ambrosio, G.~Giudice, G.~Isidori and A.~Strumia,
\newblock Nucl.Phys. {\bf B645}, 155 (2002), [hep-ph/0207036].

\bibitem{Cirigliano:2005ck}
V.~Cirigliano, B.~Grinstein, G.~Isidori and M.~B. Wise,
\newblock Nucl.Phys. {\bf B728}, 121 (2005), [hep-ph/0507001].

\bibitem{Hirsch:2012ax}
M.~Hirsch, F.~Staub and A.~Vicente,
\newblock Phys.Rev. {\bf D85}, 113013 (2012), [1202.1825].

\bibitem{Abada:2012cq}
A.~Abada, D.~Das, A.~Vicente and C.~Weiland,
\newblock JHEP {\bf 1209}, 015 (2012), [1206.6497].

\bibitem{Deppisch:2005rv}
F.~Deppisch, H.~P{\"a}s, A.~Redelbach and R.~R{\"u}ckl,
\newblock Phys.Rev. {\bf D73}, 033004 (2006), [hep-ph/0511062].

\bibitem{Ding:2009gh}
G.-J. Ding and J.-F. Liu,
\newblock JHEP {\bf 1005}, 029 (2010), [0911.4799].

\bibitem{Ishimori:2010su}
H.~Ishimori and M.~Tanimoto,
\newblock Prog.Theor.Phys. {\bf 125}, 653 (2011), [1012.2232].

\bibitem{Hirsch:2003dr}
M.~Hirsch, J.~Romao, S.~Skadhauge, J.~Valle and A.~Villanova~del Moral,
\newblock Phys.Rev. {\bf D69}, 093006 (2004), [hep-ph/0312265].

\bibitem{Deppisch:2010sv}
F.~F. Deppisch, F.~Plentinger and G.~Seidl,
\newblock JHEP {\bf 1101}, 004 (2011), [1011.1404].

\bibitem{battaglia:2001zp}
M.~Battaglia {\em et~al.},
\newblock Eur. Phys. J. {\bf C22}, 535 (2001), [hep-ph/0106204].

\bibitem{Bartl:2005yy}
A.~Bartl {\em et~al.},
\newblock Eur. Phys. J. {\bf C46}, 783 (2006), [hep-ph/0510074].

\bibitem{Carquin:2008gv}
E.~Carquin, J.~Ellis, M.~Gomez, S.~Lola and J.~Rodriguez-Quintero,
\newblock JHEP {\bf 0905}, 026 (2009), [0812.4243].

\bibitem{Wyler:1982dd}
D.~Wyler and L.~Wolfenstein,
\newblock Nucl.Phys. {\bf B218}, 205 (1983).

\bibitem{Bernabeu:1987gr}
J.~Bernabeu, A.~Santamaria, J.~Vidal, A.~Mendez and J.~Valle,
\newblock Phys.Lett. {\bf B187}, 303 (1987).

\bibitem{Ilakovac:1994kj}
A.~Ilakovac and A.~Pilaftsis,
\newblock Nucl.Phys. {\bf B437}, 491 (1995), [hep-ph/9403398].

\bibitem{Tommasini:1995ii}
D.~Tommasini, G.~Barenboim, J.~Bernabeu and C.~Jarlskog,
\newblock Nucl.Phys. {\bf B444}, 451 (1995), [hep-ph/9503228].

\bibitem{Pilaftsis:2004xx}
A.~Pilaftsis,
\newblock Phys.Rev.Lett. {\bf 95}, 081602 (2005), [hep-ph/0408103].

\bibitem{Pilaftsis:2005rv}
A.~Pilaftsis and T.~E. Underwood,
\newblock Phys.Rev. {\bf D72}, 113001 (2005), [hep-ph/0506107].

\bibitem{Kersten:2007vk}
J.~Kersten and A.~Y. Smirnov,
\newblock Phys.Rev. {\bf D76}, 073005 (2007), [0705.3221].

\bibitem{Deppisch:2010fr}
F.~F. Deppisch and A.~Pilaftsis,
\newblock Phys.Rev. {\bf D83}, 076007 (2011), [1012.1834].

\bibitem{Deppisch:2004fa}
F.~Deppisch and J.~Valle,
\newblock Phys.Rev. {\bf D72}, 036001 (2005), [hep-ph/0406040].

\bibitem{Das:2012ii}
S.~Das, F.~Deppisch, O.~Kittel and J.~Valle,
\newblock Phys.Rev. {\bf D86}, 055006 (2012), [1206.0256].

\bibitem{Pati:1974yy}
J.~C. Pati and A.~Salam,
\newblock Phys. Rev. {\bf D10}, 275 (1974).

\bibitem{Mohapatra:1974gc}
R.~Mohapatra and J.~C. Pati,
\newblock Phys.Rev. {\bf D11}, 2558 (1975).

\bibitem{Senjanovic:1975rk}
G.~Senjanovic and R.~N. Mohapatra,
\newblock Phys.Rev. {\bf D12}, 1502 (1975).

\bibitem{Duka:1999uc}
P.~Duka, J.~Gluza and M.~Zralek,
\newblock Annals Phys. {\bf 280}, 336 (2000), [hep-ph/9910279].

\bibitem{Cirigliano:2004mv}
V.~Cirigliano, A.~Kurylov, M.~Ramsey-Musolf and P.~Vogel,
\newblock Phys.Rev. {\bf D70}, 075007 (2004), [hep-ph/0404233].

\bibitem{Deppisch:2003wt}
F.~Deppisch, H.~P{\"a}s, A.~Redelbach, R.~R{\"u}ckl and Y.~Shimizu,
\newblock Phys.Rev. {\bf D69}, 054014 (2004), [hep-ph/0310053].

\end{thebibliography}

\end{document}